\documentclass[twocolumn, 10pt, pre, aps, showpacs, reprint, amsmath, amssymb, superscriptaddress, nofootinbib]{revtex4-2}

\usepackage[]{graphicx} 
\usepackage{xfrac}
\usepackage{hyperref}

\begin{document}

\newcommand{\SNR}{\mathrm{SNR}}
\newcommand{\var}{\mathrm{Var}}
\newcommand{\cov}{\mathrm{Cov}}
\newcommand{\LRF}{F_\mathrm{red}}

\title{Spontaneous phase locking in a broad-area semiconductor laser}
\author{Stefan Bittner}
\email{stefan.bittner@univ-lorraine.fr}
\author{Marc Sciamanna}
\affiliation{Laboratoire Mat\'eriaux Optiques, Photonique et Syst\`emes (LMOPS), CentraleSup\'elec, Universit\'e de Lorraine, 2 rue Edouard Belin, 57070 Metz, France}
\affiliation{Chair in Photonics, LMOPS, CentraleSup\'elec, 2 rue Edouard Belin, 57070 Metz, France}

\date{\today}

\begin{abstract}
Broad-area semiconductor lasers are employed in many high-power applications, however, their spatio-temporal dynamics is complex and intrinsically unstable due to the interaction of several transverse lasing modes. A dynamical and spatio-spectral analysis with ultra-high resolution of commercial broad-area lasers reveals multiplets of phase-locked first- and second-order transverse modes that are spontaneously created by the nonlinear dynamics for a wide range of operation parameters. Phase locking between modes of different transverse order is demonstrated by comparing the linewidths of the lasing modes to that of their beat note and by a direct measurement of their phase fluctuation correlations. The spontaneous phase locking is unexpected since the overall dynamics is unstable and the system lacks any intentional feature to induce locking. This partially synchronized dynamical state with groups of synchronized and unsynchronized laser modes coexisting is similar to what is called a chimera state in networks of coupled oscillators, hence indicating that such states may exist in a wider range of systems than previously assumed. Moreover, some of the phase-locked modes do not exist on the passive-cavity level, but are created by the nonlinear dynamics, an effect not previously observed in the context of partially synchronized laser modes.
\end{abstract}

\maketitle

\section{Introduction}
The dynamics of lasers is very rich, ranging from stable emission and periodic dynamics to deterministic chaos \cite{Narducci1988, Tartwijk1998, OhtsuboBook2013, Sciamanna2015}. Moreover, spatial pattern formation and complex spatio-temporal dynamics are observed for lasers featuring several transverse modes \cite{Abraham1990, Lugiato1992}, including gas lasers \cite{Huyet1995}, fiber lasers \cite{Jauregui2013, Guo2021} and both surface-emitting \cite{Bittner2022, Giudici1998} and edge-emitting \cite{Fischer1996, Marciante1998, Scholz2008} semiconductor lasers. Such broad-area lasers (BALs) are needed to reach high output powers beyond the limitations of single transverse-mode lasers. However, they are susceptible to spatio-temporal instabilities stemming from the interactions between transverse lasing modes mediated by the active medium. While progress has been made in controlling spatio-temporal dynamics with special device designs \cite{Wright2017, Wright2020, Bittner2018a, Kim2023b, Ivars2023, Chen2023d}, the documentation and understanding of dynamical phenomena occurring in simple broad-area lasers remains fragmentary. 

Beyond application-related considerations, BALs are hence interesting model systems for studying complex multimode dynamics. Important issues are the possibly chaotic nature of the spatio-temporal dynamics, how lasing modes organize in space, wavelength and time, and if synchronization or other forms of collective dynamics of lasing modes occur. An interesting form of collective dynamics exhibiting partial synchronization are chimera states \cite{Abrams2004, Tinsley2012, Parastesh2021} in which groups of synchronized oscillators coexist with unsynchronized ones. Chimera states have been predicted numerically for arrays of coupled lasers \cite{Shena2017} and demonstrated experimentally for semiconductor lasers with feedback \cite{Larger2015, Uy2019} and absorptive sections \cite{Viktorov2014}. Similarly, synchronization of mode clusters was found for a quantum cascade laser \cite{Kazakov2024}. However, previous works concerned only lasers with a single transverse mode, whereas cluster synchronization of different transverse modes has not been studied before. 

We investigate a broad-area edge-emitting semiconductor laser with a quantum well as active medium. Semiconductor BALs suffer from intrinsic spatio-temporal instabilities caused by the interplay of spatial-hole burning, carrier-induced index changes and diffraction \cite{Hess1994, Fischer1996, Hess1996, Adachihara1993, Marciante1997, Marciante1998}. Their spatio-temporal dynamics displays significant long-range correlations \cite{Hess1994, Fischer1996, Arahata2015, Tachikawa2010} which has been attributed to partial self-mode locking \cite{Hess1996, Kaiser2004, Kaiser2006a}. While mode-locking between longitudinal and transverse modes can be induced by pulsed optical injection \cite{Kaiser2004, Kaiser2006a}, phase- or mode-locking effects in conventional, free-running quantum-well BALs have never been experimentally demonstrated. 

We perform dynamical and spatio-spectral measurements in order to identify the different transverse modes and find possible correlations between them. We observe an undocumented dynamical effect, the creation of multiplets of 1st- and 2nd-order phase-locked transverse modes that coexist with unlocked modes. This partially synchronized state is observed for two different lasers and in a wide range of operation parameters. 

\section{Experimental results} \label{sec:experiments}

\begin{figure}[tb]
\begin{center}
\includegraphics[width = 8.4 cm]{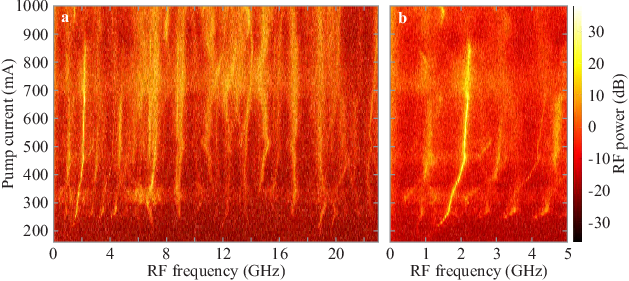}
\end{center}
\caption{\textbf{Emission properties of the BAL.} \textbf{a},~RF-spectra of the total laser emission and \textbf{b},~magnification around the particularly narrow frequency component near $2$~GHz.}
\label{fig:RFspectra}
\end{figure}

We investigate a commercial broad-area laser (see Appendix~\ref{ssec:BAL}) with continuous wave pumping in solitary operation (no optical injection, feedback or modulation). The lateral confinement in the $50~\mu$m wide and $1.5$~mm long cavity is due to a combination of gain and index guiding, and the laser threshold is $I_{th} = 160$~mA (see Fig.~\ref{sfig:LIcurve}). 

Figure~\ref{fig:RFspectra}a shows the RF-spectra of the BAL as function of the pump current. They were measured by coupling the entire laser beam into a multimode fiber (MMF) connected to a fast photodetector (see Appendix~\ref{ssec:timeDomMeas}). The laser shows instabilities, and the RF-spectra feature a number of discrete and partially overlapping peaks on a broad background. Their typical linewidths (FWHM) are $100$~MHz or more. However, one peak around $2$~GHz is significantly narrower (see Fig.~\ref{fig:RFspectra}b) with a linewidth below $10$~MHz for pump currents up to $600$~mA. It becomes broader and disappears as the pump current increases further (see Fig.~\ref{sfig:thePeakProperties}). We surmise that the peaks in the RF-spectra are beat notes between different lasing modes \cite{Stelmakh2014}, in which case the narrow peak could be explained by the beating of phase-locked modes. 

In order to confirm this, we perform spatio-spectral measurements, but the resolution of conventional imaging spectrometers is insufficient (see Appendix~\ref{sec:lowResSpatioSpec}). Sophisticated grating-based setups enable spatio-spectral measurements with resolutions as good as $120$~MHz \cite{Stelmakh2006, Crump2012, Crump2014, Uhlig2023}, which is sufficient to resolve the different transverse modes. However, in order to detect phase locking effects, we also need to measure the linewidths of the lasing modes which are of the order of $100$~MHz for multimode semiconductor lasers \cite{Elsasser1985}. Hence, we adopt a heterodyning technique with even better spectral resolution \cite{Brunner2015, Wishon2018, Stelmakh2014} that additionally permits measuring the phases of lasing modes \cite{Cappelli2019}. 

\begin{figure*}[tb]
\begin{center}
\includegraphics[width = 17 cm]{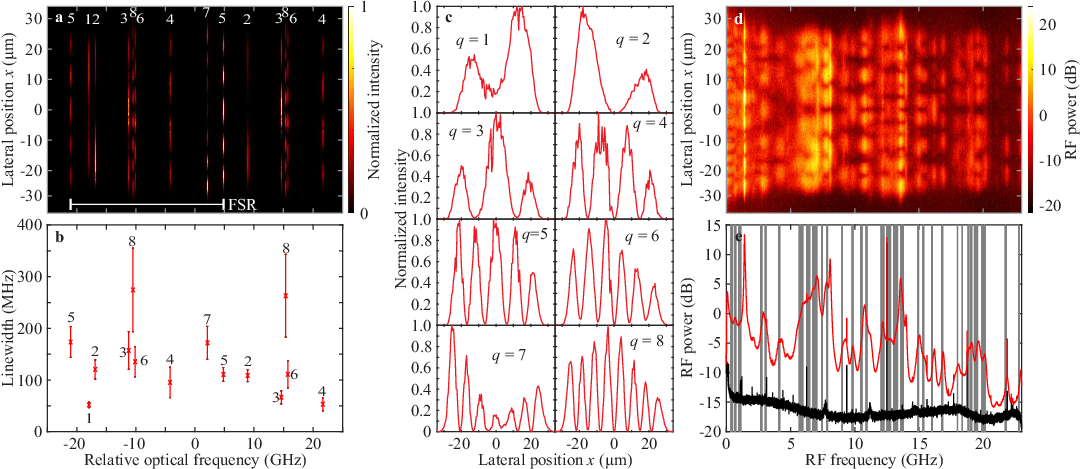}
\end{center}
\caption{\textbf{Heterodyne measurement results at 1000~mA.} \textbf{a},~Reconstructed spatio-spectral image around a reference wavelength of $851.142$~nm. The white numbers indicate the transverse quantum numbers $q$ of the modes, and the white bar the free spectral range. All modes are plotted with unit amplitude. \textbf{b},~Linewidths of the lasing modes. \textbf{c},~Typical mode profiles for modes with transverse quantum numbers $q = 1$--$8$. \textbf{d},~Spatially resolved RF-spectrum. \textbf{e},~Spatially-averaged RF-spectrum (red) and detector background signal (black). The vertical gray lines indicate the beat frequencies of the lasing modes.}
\label{fig:caseHighPump}
\end{figure*}

For spatially-resolved heterodyne measurements, we collect the BAL emission in the image plane of its output facet with a MMF whose lateral position is scanned (see Fig.~\ref{sfig:setupHetDyne}). The signal of the BAL interferes with a single-mode reference laser on a fast photodetector to create beat notes from which we reconstruct the optical spectrum of the BAL (see Appendix~\ref{sec:spatHetdyneMeas}). The spectral resolution of the heterodyne measurement is fundamentally limited by the linewidth of the reference laser ($< 1$~MHz). 

We first consider a pump current of $1000$~mA for which there is no narrow peak in the RF-spectrum. The reconstructed spatio-spectral image (see Appendix~\ref{ssec:hetDyneAna}) in Fig.~\ref{fig:caseHighPump} covers almost two longitudinal mode spacings (free spectral range, FSR, $25.9$~GHz) and features transverse modes of order $q = 1$--$8$, where $q = 1$ is the fundamental mode. Since the spacing between the 1st-order transverse mode and those with $q = 6$--$8$ is larger than the FSR, different longitudinal-mode groups mix, and the modes with $q = 3$, $6$ and $8$ are spectrally close as a consequence. The linewidths of the lasing modes (Fig.~\ref{fig:caseHighPump}b) are in the range of $50$--$300$~MHz. 

Figure~\ref{fig:caseHighPump}c presents typical mode profiles, which show significant deviations from those expected in a passive Fabry-Perot (FP) cavity with homogeneous refractive index distribution. First, the fundamental mode $q = 1$ has a dip in the center and thus two peaks instead of one, in contrast to observations for other devices \cite{Crump2012, Mandre2005, Stelmakh2006}. Second, the peak heights are not uniform. For example, the $q = 3$ and $8$ modes have higher peaks in the center, whereas the $q = 7$ mode has lower peaks in the center. Third, several modes have asymmetries even though the spectrally-integrated near-field intensity distributions are symmetric (see Fig.~\ref{sfig:nf-ff-imgs}). In particular, the right (left) peak of the 1st (2nd) order mode is always higher. Thus, these two modes have complementary profiles as if they were competing for gain. It should be noted that the features discussed above are systematically found in all measurements. In a cavity with only index guiding in the lateral direction, the peak heights would of course be homogeneous \cite{Crump2012}, but an inhomogeneous current density as well as thermal lensing effects can significantly modify the mode profiles. In Ref.~\cite{Crump2012}, thermal lensing was found to be the dominant mechanism, though no explanation for the observed asymmetries was given. Dynamical effects like spatial and spectral hole burning can play an important role as well. 

The spatially-resolved RF-spectrum in Fig.~\ref{fig:caseHighPump}d shows that different frequency components exhibit different spatial patterns, which is expected for beating of different transverse modes \cite{Stelmakh2014}. Figure~\ref{fig:caseHighPump}e compares the spatially-averaged RF-spectrum to the beat notes (i.e., optical frequency differences) of the lasing modes obtained from six heterodyne measurements at $1000$~mA. Almost all major peaks and frequency components in the RF-spectrum can thus be explained as beat notes. The widths of the peaks in the RF-spectrum are typically $100$~MHz and more and thus consistent with the widths of the lasing modes (see Fig.~\ref{fig:caseHighPump}b). However, the significant offset of the RF-spectrum compared to the background signal over the whole frequency range and the few unexplained peaks indicate that dynamical effects beyond mode beating contribute as well. 

\begin{figure*}[tb]
\begin{center}
\includegraphics[width = 17 cm]{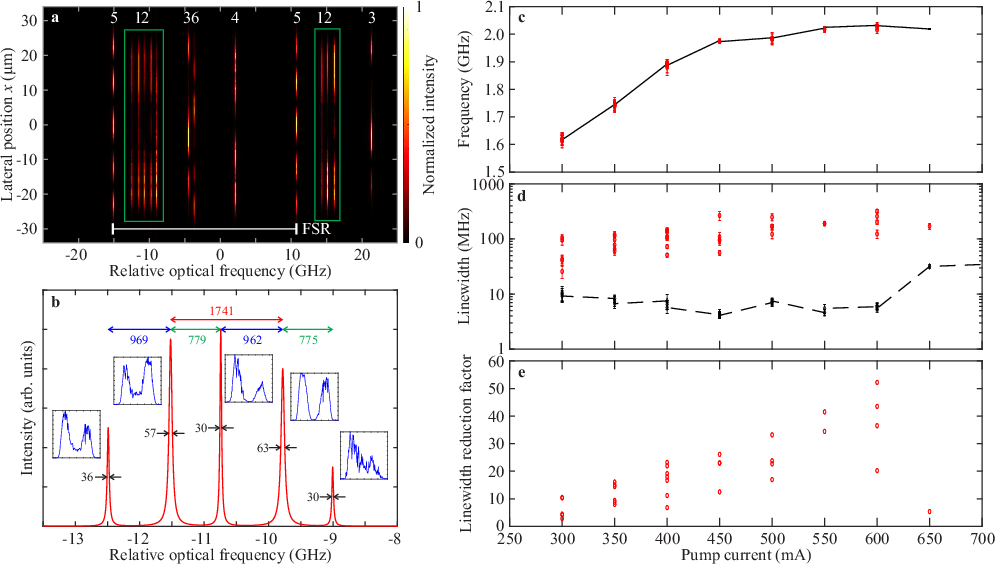}
\end{center}
\caption{\textbf{Heterodyne measurement results at 350~mA.} \textbf{a},~Reconstructed spatio-spectral image around a reference wavelength of $849.594$~nm. The green boxes mark a quintuplet and a triplet of 1st- and 2nd-order modes. All modes are plotted with unit amplitude. \textbf{b},~Schematic of the quintuplet. The insets show the spatial profiles of the different modes, and the black arrows indicate their linewidths. The spacings and linewidths are given in MHz. \textbf{c},~Frequency of the narrow peak in the RF-spectrum (black) and NNN-spacings of multiplets (red) as function of the pump current. \textbf{d},~Linewidth of the narrow peak (black) and sum of linewidths of NNN-mode pairs (red). \textbf{e},~Linewidth reduction factor for different NNN-mode pairs, which indicates the local degree of synchronization.}
\label{fig:caseMultiplet}
\end{figure*}

We next consider lower pump currents for which the narrow peak in the RF-spectra is observed. The spatio-spectral image for $350$~mA in Fig.~\ref{fig:caseMultiplet}a shows transverse modes up to order $q =6$. Surprisingly, we observe a quintuplet and a triplet of 1st- and 2nd-order modes (indicated by the green boxes), that is, additional lasing modes are created by the nonlinear dynamics. The schematic of the quintuplet in Fig.~\ref{fig:caseMultiplet}b shows two 1st-order modes followed by two 2nd-order modes and a fifth mode whose spatial profile does not allow a clear identification. Many such multiplets are found for $300$ to $650$~mA, always consisting of modes with $q = 1$ and $2$, but also longitudinal mode groups with only one transverse mode of each order like for higher pump currents are observed. 

Interestingly, the nearest-neighbor (NN) spacings indicated in Fig.~\ref{fig:caseMultiplet}b are not equal, instead, every second spacing is different. This means that four-wave mixing cannot be responsible for creating the multiplets since it would result in equidistant modes. More importantly, the next-nearest-neighbor (NNN) spacings are all the same and equal to the frequency of the narrow peak in the RF-spectrum shown in Fig.~\ref{fig:RFspectra}b. So this peak results from to the beating of pairs of modes with different transverse order of which one was created by nonlinear dynamics. Figure~\ref{fig:caseMultiplet}c shows perfect agreement between the frequency of the narrow RF-peak and the NNN spacings in the current range of interest. 

The frequency of this narrow beat note has a complicated dependency on pump current and temperature (see Fig.~\ref{sfig:thePeakProperties}). It increases significantly with the pump current, but does not exhibit the square-root scaling of the relaxation oscillation frequency. It is of the same order as two times the NN spacing between 1st- and 2nd-order modes which increases with pump current (see Fig.~\ref{sfig:NNspacings}). 

The linewidths of the lasing modes in the quintuplet are indicated by the double arrows in Fig.~\ref{fig:caseMultiplet}b. Every second mode has a clearly larger linewidth than its neighbors (see also Fig.~\ref{sfig:multipletLinewidths}). More importantly, the mode linewidths are significantly larger than that of the beat note, which is smaller than $10$~MHz in the regime up to $600$~mA (see Fig.~\ref{fig:caseMultiplet}d). If the phase fluctuations of the lasing modes were independent, the linewidth of their beat note would be equal to the sum of their linewidths (see Section~\ref{sssec:linewidthTheory}). Instead, the sum of the linewidths of the NNN-mode pairs is of the order of $100$~MHz in most cases (Fig.~\ref{fig:caseMultiplet}d). So the linewidth of the beat note is typically reduced by a factor (called linewidth reduction factor, LRF) ranging from $10$ to $50$ compared to the sum of the mode linewidths as shown in Fig.~\ref{fig:caseMultiplet}e. This means that the phase fluctuations of the NNN-mode pairs are highly correlated, i.e., these lasing modes are phase-locked. This spontaneous phase locking is surprising since the BAL is in solitary operation without feedback, injection \cite{Kaiser2004, Kaiser2006a}, modulation or absorbing sections \cite{Viktorov2014, Kazakov2024}. It should be noted that the LRF shown in Fig.~\ref{fig:caseMultiplet}e increases till $600$~mA, i.e., the phase locking gets stronger, until it is lost around $650$~mA where the beat note becomes broader again (see also Fig.~\ref{sfig:thePeakProperties}). 

The analysis of the linewidths is not a proof of phase locking of particular mode pairs since (i) it is not clear which NNN-mode pairs in the multiplets are phase-locked or not and (ii) we cannot exclude the existence of mode pairs with narrow linewidths in other longitudinal mode groups that could explain the narrow beat note without phase locking. Furthermore we wonder if the NN-mode pairs may be weakly phase-locked even though their beat note is not particularly narrow. Fortunately we can extract the phase fluctuation correlations (PFCs) between different modes from the time-domain measurements of the heterodyne signals. 

\begin{figure*}[tb]
\begin{center}
\includegraphics[width = 17 cm]{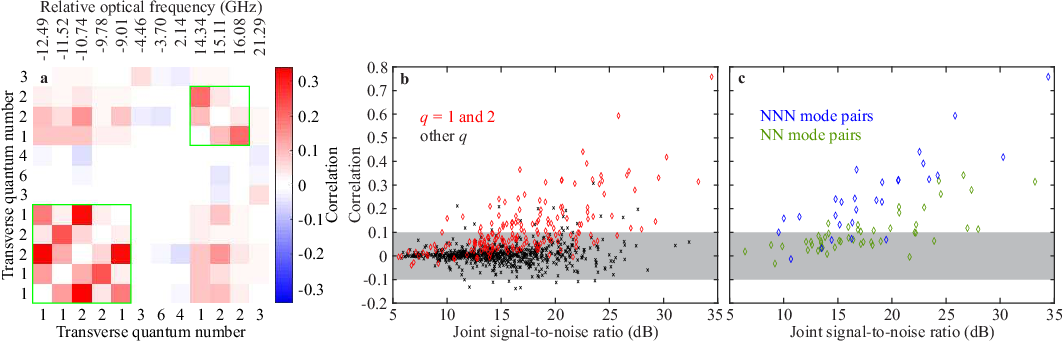}
\end{center}
\caption{\textbf{Phase fluctuation correlations.} \textbf{a},~Matrix of phase fluctuation correlations for the measurement shown in Fig.~\ref{fig:caseMultiplet}a. The green boxes indicate the quintuplet and the triplet. \textbf{b},~Phase fluctuation correlations as function of the joint signal-to-noise ratio for all mode pairs in data sets with multiplets. Pairs of first- and second-order modes are indicated in red, all other pairs in black. The gray rectangle indicates the range of not significant correlations, $[-0.1, +0.1]$. \textbf{c},~Phase fluctuation correlations of NN-mode pairs (green) and NNN-mode pairs (blue) only.}
\label{fig:phaseCorr}
\end{figure*}

The calculation of the phase fluctuation correlations is explained in Appendix~\ref{sec:phaseFlucCorr}. Figure~\ref{fig:phaseCorr}(a) shows the PFC matrix for the lasing modes presented in Fig.~\ref{fig:caseMultiplet}, with the quintuplet and triplet marked by green boxes. Most mode pairs exhibit negligibly small correlations, whereas significant correlations are observed between the modes of the multiplets. The highest correlations are found for NNN-mode pairs (second off-diagonal), all of which are clearly correlated. There are also correlations between the NN-mode pairs (first off-diagonal), but they are weaker. Moreover, it should be noted that some correlations are found between the modes of the quintuplet and the triplet which are one FSR away. These findings are exemplary for all the multiplets and are consistent with the linewidths of lasing modes and their beat notes as discussed above. They clearly prove that the modes created by the nonlinear dynamics are phase-locked to their next-nearest neighbors, which suggests that a phase-sensitive nonlinear process is responsible for their creation. 

However, the measured PFCs are lower than expected from the linewidth reduction factors shown in Fig.~\ref{fig:caseMultiplet}e. For example, a LRF of $10$ corresponds to a PFC of $0.9$ (see Fig.~\ref{sfig:LRFvsCorr}), but such high values are not observed. In fact, the calculation of the PFCs is very sensitive to perturbations by measurement noise and interfering signals from the BAL dynamics. Figure~\ref{fig:phaseCorr}b shows the PFCs of all measurements featuring multiplets as function of the joint signal-to-noise ratio (SNR), which is the mean of the SNRs of the two heterodyne signals (see Appendix~\ref{sssec:phaseFlucExpAna} for details). We see a clear increase of the maximal observed PFCs with the joint SNR. Hence, one should compare the PFC of a mode pair to that of other mode pairs with the same joint SNR in order to determine if the PFC can be considered high or not. 

Most mode pairs in Fig.~\ref{fig:phaseCorr} have PFCs in the range of $[-0.1, +0.1]$, which we consider as not significant. The comparison of correlations between modes of 1st and 2nd order [i.e., pairs with transverse quantum numbers $(1, 1)$, $(1, 2)$ or $(2, 2)$] to all other mode pairs shows that only the modes with $q = 1$ and $2$ exhibit significant correlations, with a few exceptions to be discussed elsewhere. Figure~\ref{fig:phaseCorr}c shows only the PFCs of NN- and NNN-mode pairs in multiplets, where the NNN-correlations are systematically higher than the NN-correlations for a given joint SNR. This confirms that the phase locking between NNN-modes is stronger compared to NN-pairs. 

\section{Discussion}
Our experiments reveal a spontaneous phase locking of transverse modes in a simple broad-area Fabry-Perot laser without special features or external control elements. This is surprising given the significant spatio-temporal instabilities of such devices. Since only two out of up to eight transverse mode orders exhibit phase locking, the dynamics of the BAL remains complex and unstable. Hence the phase-locking effect is not easily recognized, and direct measurements of the phase fluctuation correlations are needed to ascertain which modes are locked. 

Mode locking can be generated by techniques like saturable absorbers, external modulation and feedback, and pulsed injection into broad-area lasers can induce longitudinal \cite{Kaiser2004} and transverse mode-locking \cite{Kaiser2006a}. Furthermore, spatio-temporal mode locking has been demonstrated for Helium-Neon lasers \cite{Smith1968}, Ti:Sapphire lasers \cite{Cote1998} and fiber lasers \cite{Wright2017, Wright2020}, but requires careful design of the laser system and its cavity. Here, in contrast, phase-locked multiplets emerge \textit{spontaneously} from the intrinsic laser dynamics of a BAL with a normal Fabry-Perot cavity. While spontaneous phase, frequency and mode locking have been previously observed for semiconductor lasers, our case is different in several respects. 

For semiconductor lasers with quite narrow FP cavities featuring only the 1st- and 2nd-order transverse modes, phase locking between the transverse modes has been observed \cite{Ziegler1999}, which leads to a periodic oscillation of the beam profile, as well as frequency locking \cite{Tan1997, Fu1998}, which results in asymmetric, static emission profiles. Phase locking \cite{Choi2008a, Weng2017} and frequency locking \cite{Harayama2003a, Fukushima2005} between two different spatial modes have also been observed for asymmetric cavities. However, in these examples only two different spatial modes lase in the first place, whereas our BAL features up to eight transverse lasing modes. Moreover, previously only modes that naturally exist in the passive cavity were locked, whereas we observe multiplets of modes created by the nonlinear dynamics. 

Spontaneous mode locking between longitudinal modes has been observed for quantum-dot \cite{Rosales2012a}, quantum-dash \cite{Rosales2012b} and quantum-cascade lasers (QCLs) \cite{Hugi2012, Burghoff2014}. Furthermore, QCL frequency combs with phase and frequency locking of several transverse modes have been demonstrated \cite{Yu2009, Wojcik2011}. However, the dynamics in these lasers is quite different from that of quantum well lasers due to the much shorter gain recovery time \cite{Gioannini2015} and the strong third-order nonlinearity of the former \cite{Hugi2012}. Locking of transverse modes in QCLs is primarily attributed to four-wave mixing \cite{Yu2009, Wojcik2011}, whereas our observations cannot be explained by four-wave mixing since the multiplet modes are not equidistant. 

\begin{figure}[tb]
\begin{center}
\includegraphics[width = 8.4 cm]{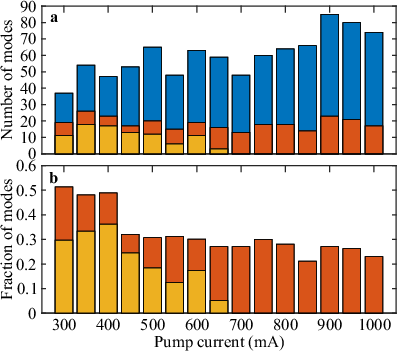}
\end{center}
\caption{\textbf{Global degree of synchronization.} \textbf{a},~Total number of measured modes (blue), number of modes with transverse quantum numbers $q = 1$ and $2$ (red), and number of phase-locked modes in multiplets (yellow). \textbf{b},~Fraction of modes with transverse quantum numbers $q = 1$ and $2$ (red) and of phase-locked modes in multiplets (yellow). The fraction of phase-locked modes can be considered as an order parameter indicating the degree of global synchronization. In contrast, the LRF in Fig.~\ref{fig:caseMultiplet}e represents the degree of local synchronization, which shows a qualitatively different dependency on the pump current.}
\label{fig:synchDeg}
\end{figure}

The most interesting aspect of our laser is the coexistence of groups of coherent modes (the phase-locked multiplets) with a large number of incoherent (unlocked) modes, similar to what is called chimera state in networks of coupled oscillators \cite{Abrams2004, Tinsley2012, Parastesh2021}. Figure~\ref{fig:synchDeg} compares the total number of modes measured (blue) for each pump current with the number of 1st- and 2nd-order modes (red) and those belonging to multiplets (yellow). Even though the total number of measured modes is modest, we observe a clear trend as function of the current: the fraction of 1st- and 2nd-order modes decreases above $400$~mA as more higher-order transverse modes are excited, but remains around $20$--$30\%$. In contrast, the fraction of phase-locked modes, which is an order parameter indicating the global degree of synchronization \cite{Shena2017}, increases up to $400$~mA, then decreases, and vanishes completely above $650$~mA. This abrupt emergence of partial order at a certain parameter threshold is characteristic of the transition to chimera states in networks of coupled oscillators~\cite{Kuramoto1984a, Shena2017}. 

Chimera states have been reported for semiconductor lasers with feedback \cite{Larger2015, Uy2019}, where virtual oscillators in the feedback loop are considered. Cluster synchronization were furthermore found in quantum-dot \cite{Viktorov2014} and quantum-cascade \cite{Kazakov2024} lasers with absorbing sections. Our system is different in several respects. Partial synchronization is not induced by feedback or absorbing sections, but results from the inherent laser dynamics, and is formed between different transverse modes. In contrast to Refs.~\cite{Larger2015, Uy2019}, we observe synchronization between actual lasing modes instead of virtual oscillators. Moreover, some of the synchronized lasing modes do not exist in the system \textit{a priori}, but are created by the nonlinear dynamics, another distinction from previous works. 

Our observation of groups of phase-locked modes in a commercial semiconductor laser indicates that partially synchronized dynamics may be more frequent in lasers and other nonlinear systems than previously thought. The spatio-temporal dynamics of broad-area lasers is still not fully understood, and our observation of spontaneous phase locking gives a new perspective for understanding and possibly controlling their instabilities using, e.g., optical injection or special cavity designs. Figure~\ref{fig:synchDeg} shows that the pump current determines the degree of synchronization, but further studies are needed to understand the underlying physical mechanisms. This will have important implications for our fundamental understanding of complex spatio-temporal and multimode dynamics as well as for applications like spatio-temporal mode locking and high-power lasers. 

\acknowledgments{The Chair in Photonics is supported by R\'egion Grand Est, GDI Simulation, D\'epartement de la Moselle, European Regional Development Fund, CentraleSup\'elec, Fondation CentraleSup\'elec, and the Eurometropole de Metz. S.B.\ thanks Mario Fernandes for technical support.}

\newpage
\appendix

\section{Experimental setup and basic measurements}

\subsection{Broad-area laser} \label{ssec:BAL}
We use a commercial quantum-well semiconductor laser (Sheaumann Laser, P/N CM-850-1000-050) soldered to a C-mount. The cavity is laterally defined by etching to just above the active layer and depositing a dielectric afterwards, which gives the laser an aspect of both gain and index guiding. The laser is operated in a temperature-controlled laser diode mount (Thorlabs LDMC20/M) and pumped in continuous wave operation by a diode driver (Thorlabs ITC4005). The temperature of the mount is kept constant at $21^\circ$C during all measurements unless otherwise noted. The laser emission is transverse-electric polarized (electric field parallel to the cavity plane). 

\begin{figure}[b]
\begin{center}
\includegraphics[width = 8.4 cm]{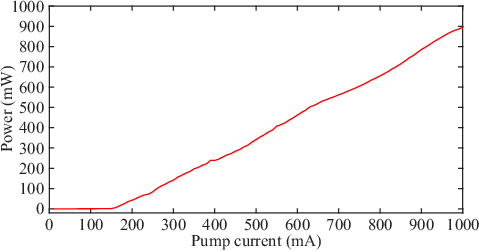}
\end{center}
\caption{\textbf{LI-curve of the broad-area laser.} The LI-curve was measured in free space and shows a threshold at $160$~mA.}
\label{sfig:LIcurve}
\end{figure}

Figure~\ref{sfig:LIcurve} shows the LI-curve of the broad-area laser. It was measured in free space by using an objective and a lens to collect and focus the laser emission on a power meter (Newport 818-IG). The threshold is at $I_{th} = 160$~mA. The maximal measured power is in good agreement with the specifications ($1$~W at $1100$~mA).

\subsection{Near- and far-field intensity distributions}

\begin{figure*}[tb]
\begin{center}
\includegraphics[width = 14 cm]{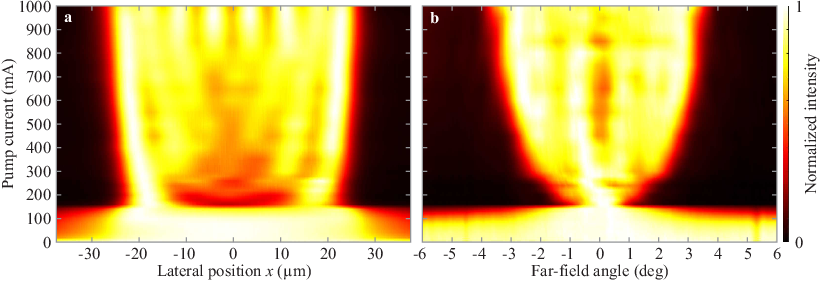}
\end{center}
\caption{\textbf{Near- and far-field intensity distributions.} \textbf{a}, Near-field images. \textbf{b}, Far-field images. The intensity distributions for each pump current are normalized to unit amplitude.}
\label{sfig:nf-ff-imgs}
\end{figure*}

The output facet of the BAL is imaged onto a CMOS camera (Allied Vision Mako U-503B) using a $25\times$ microscope objective (NA $= 0.5$) and a lens with $175$~mm focal length (Thorlabs LA1399-B) in a telescope configuration. A beam splitter and a third lens (Thorlabs LA1050-B, $100$~mm focal length) are used to obtain the Fourier transform of the image plane for parallel far-field measurements with a second camera (Allied Vision Mako U-130B). The far-field angle is calculated from the magnification of the telescope and the focal length of the third lens. Neutral-density filters in the beam path and the exposure times of the cameras are adjusted during the measurement to avoid both low signal strength and saturation. 

Figure~\ref{sfig:nf-ff-imgs}a shows the measured near-field intensity distributions, which are almost uniformly flat well below threshold ($I_{th} = 160$~mA). Just above threshold, the intensity distribution exhibits a strong dip in the center with two peaks near the edges of the output facet. As the pump current increases, these two peaks move further towards the edges, and the center of the intensity distribution slowly fills. Above approximately $850$~mA the near-field distribution develops a top hat profile with significant modulation (cf.\ Fig.~\ref{sfig:spatioSpecImg}). A similar evolution of the near-field intensity distributions as function of the pump current was observed in Ref.~\cite{Asatsuma2006} and attributed to the interplay of index guiding and carrier-induced index changes. It should be noted that the near-field images are very symmetric around $x = 0$ even though the individual transverse mode profiles show significant asymmetries (see Fig.~\ref{fig:caseHighPump}c.). 

The far-field intensity distributions shown in Fig.~\ref{sfig:nf-ff-imgs}b are very broad below threshold and collapse to a single lobe at $0^\circ$ just above threshold. In this current regime, mainly the 1st- and 2nd-order modes lase, but the near-field distributions exhibit a pronounced dip in the center nonetheless since the fundamental mode features a dip in the center similar to the $q = 2$ mode (cf.\ Fig.~\ref{fig:caseHighPump}c). As the pump current increases, higher-order modes start to lase, and the power shifts away from the low order modes. 

\subsection{Time-domain measurements} \label{ssec:timeDomMeas} 
Time-domain measurements were performed with a fast photodetector (Newport 1484-A-50, $15$~kHz to $22$~GHz) connected to a traveling-wave amplifier (Newport 1421, $8.5$~dB) and an oscilloscope (Tektronix DPO72304SX, $23$~GHz). Time traces were measured with $50$~GSample/s ($20$~ps sampling step) and $10^6$ points ($20~\mu$s long). RF-spectra $P_{RF}(f)$ are obtained by calculating the modulus squared of the Fourier transform $\mathcal{F}$ of the time traces $U(t)$ and smoothing with a moving average,
\begin{equation} \label{eq:RFspec} P_{RF}(f) = \mathcal{M} \{ |\mathcal{F}[U(t)]|^2 \} \, , \end{equation}
where $\mathcal{M}$ is the moving average over width $f_\mathrm{avg}$. A moving average width of $f_\mathrm{avg} = 2$~MHz was used unless otherwise noted. 

It should be noted that extremely narrow peaks are observed in many RF-spectra: these are artifacts due to synchronization issues between different analog-to-digital converters of the oscilloscope and appear even without input signal (see background signal in Fig.~\ref{fig:caseHighPump}e). Hence they have no physical significance and can be ignored. 

For measurements of the total laser emission, the entire beam was coupled into a graded-index multimode fiber (MMF, Thorlabs M116L02, $50~\mu$m diameter, NA$ = 0.20$, $2$~m long) using a $40\times$ aspheric objective lens (Newport 5722-B-H, NA = $0.6$) and a fiber coupler (Thorlabs MBT613D/M). Feedback from reflections at the fiber facet were avoided by means of an optical isolator (Thorlabs IO-5-850-VLP). The multimode fiber was connected to the photodetector. It should be noted that coupling the beam of a multi-transverse mode laser into a single-mode fiber is very inefficient, and more importantly the higher-order transverse modes have a lower coupling efficiency compared to low-order modes, which results in an observational bias (see Appendix~\ref{sec:MMF}). 

\subsection{Low-resolution optical spectra} \label{sec:lowResSpectra}

\begin{figure*}[tb]
\begin{center}
\includegraphics[width = 16 cm]{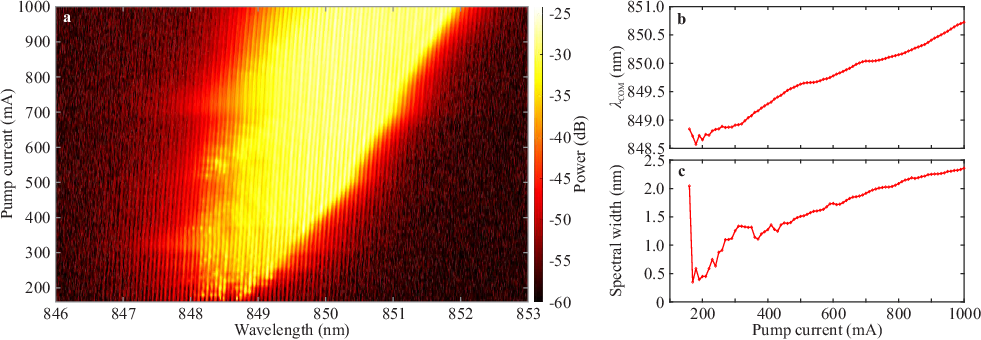}
\end{center}
\caption{\textbf{Spectra of total laser emission measured with an OSA.} \textbf{a}, Spectra of the BAL as function of the pump current. \textbf{b}, Center-of-mass wavelength $\lambda_\mathrm{COM}$. \textbf{c}, Width of spectra.}
\label{sfig:spectra-OSA}
\end{figure*}

Figure~\ref{sfig:spectra-OSA} shows lasing spectra of the BAL measured with an optical spectrum analyzer (OSA, Anritsu MS9740A) with $30$~pm ($12.5$~GHz) resolution. The entire beam was coupled into the multimode fiber (see Appendix~\ref{ssec:timeDomMeas}). While the spectral resolution of the OSA is relatively low, it allows measurements over a broader spectral range than the heterodyning technique. Due to the low resolution, the spectra feature a single series of peaks with distance equal to the longitudinal-mode spacing (free spectral range, FSR) of $25.9$~GHz ($62.4$~pm), hiding the fact that the laser exhibits several transverse modes. As the pump current increases, the center of mass (COM) of the spectrum red-shifts (see Fig.~\ref{sfig:spectra-OSA}b) due to Joule heating. The total shift is about $2$~nm over $800$~mA. The red-shift of the individual peaks in the spectrum can also be seen in Fig.~\ref{sfig:spectra-OSA}a. 

The width of the spectrum, here given by the participation ratio
\begin{equation} \left[ \int d\lambda \, P(\lambda) \right]^2 / \int d\lambda \, P^2(\lambda) \end{equation}
where $P(\lambda)$ is the optical power measured by the OSA, increases from $0.35$~nm just above threshold to $2.36$~nm at $1000$~mA, which corresponds to about $38$ longitudinal mode groups. With $8$ transverse modes, we estimate that the BAL emits in up to $300$ lasing modes in parallel, though the actual number is not clear and may be lower since not every transverse mode is found in every longitudinal mode group. 

\subsection{Low-resolution spatio-spectral images} \label{sec:lowResSpatioSpec}

\begin{figure}[tb]
\begin{center}
\includegraphics[width = 8 cm]{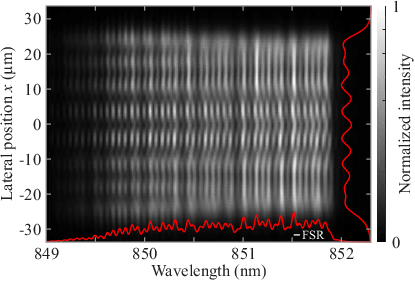}
\end{center}
\caption{\textbf{Spatio-spectral near-field image at 1000~mA measured with an imaging spectrometer}. The spatial and spectral projections are indicated in red. The small white bar indicates the free spectral range (FSR).}
\label{sfig:spatioSpecImg}
\end{figure}

Spatio-spectral images with low resolution were measured with an imaging spectrometer (Princeton Instruments SpectraPro HRS-500 with $1800$~g/mm grating, $30$~pm resolution) and a CCD camera (Allied Vision Mako U-503B). In order to measure the spectrally-resolved near-field intensity distributions, the output facet of the BAL was imaged onto the entrance slit of the spectrometer with $40\times$ magnification. A dove prism (Thorlabs PS992M-B) was used to rotate the image of the BAL facet by $90^\circ$ so it matches the vertical orientation of the entrance slit. 

The spatio-spectral image for $1000$~mA in Fig.~\ref{sfig:spatioSpecImg} shows spatial patterns varying with the wavelength due to the presence of several transverse modes. However, the individual transverse modes cannot be resolved. The spectrally-integrated near-field intensity distribution has a top-hat profile with regular modulation which indicates the presence of high-order transverse modes (see Fig.~\ref{sfig:nf-ff-imgs} for more details). The spatially integrated spectrum shows only a single series of peaks with spacing equal to the FSR like the measurements with the OSA in Fig.~\ref{sfig:spectra-OSA}, obfuscating the existence of several transverse modes. 

\section{Spatially-resolved heterodyne measurements} \label{sec:spatHetdyneMeas}

\subsection{Experimental setup and measurement protocol}

\begin{figure}[tb]
\begin{center}
\includegraphics[width = 8.4 cm]{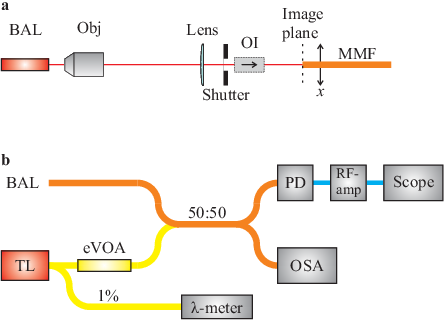}
\end{center}
\caption{\textbf{Experimental setup for spatially-resolved heterodyne measurements.} \textbf{a}, Free-space part of the setup, where $x$ indicates the lateral position of the fiber. BAL: broad-area laser, Obj: $40\times$ objective lens, OI: optical isolator, MMF: multimode fiber. \textbf{b}, Fiber-optic part of the setup. Single-mode fibers (multimode fibers) are indicated in yellow (orange), and coaxial cables in blue. TL: tunable laser (reference laser), eVOA: electronic variable optical attenuator, $\lambda$-meter: wavelength meter, PD: fast photodetector, RF-amp: RF-amplifier, OSA: optical spectrum analyzer, Scope: oscilloscope}
\label{sfig:setupHetDyne}
\end{figure}

The experimental setup is presented in Fig.~\ref{sfig:setupHetDyne}. A $40\times$ aspherical objective lens (Newport 5722-B-H, NA = $0.6$) and a plano-convex lens (Thorlabs LA1979-B, focal length $200$~mm) image the output facet of the BAL with $40\times$ magnification in the plane where its emission is collected with a MMF (Thorlabs M116L02, see above). An optical isolator (Thorlabs IO-5-850-VLP) avoids parasitic feedback into the BAL from reflections at the fiber facet. The fiber is mounted on a motorized translation stage (Thorlabs NRT100/M with driver BSC201) and scans the transverse position on the output facet of the BAL with a step size of $20~\mu$m (141 positions). The corresponding step size on the BAL output facet, $0.5~\mu$m, is slightly smaller than the spatial resolution of $0.7~\mu$m estimated from the NA of the objective lens. 

The signals from the BAL and the tunable reference laser (Toptica DL pro with Diode AR 860 nm) are superimposed via a multimode fiber-optic coupler (Thorlabs TM50R5F2B) on a fast photodetector and measured with an oscilloscope (see above). The optical powers of the reference laser and the BAL on the photodetector are adjusted to about $0.75$~mW and maximally $0.5$~mW, respectively, using a neutral density filter and a variable optical attenuator (VOA), though the power of the BAL can be smaller for low pump currents. The reference laser is set to $20^\circ$C with $143$~mA pump current and its wavelength adjusted to the desired value. It is left to stabilize for at least $20$~minutes, and we check that it is in a stable single-mode lasing regime. Its wavelength is monitored using a wavelength meter (High Finesse Angstrom WS6-600 IRVIS) connected via a $99$:$1$ fiber coupler (Thorlabs TW850R1A2). 

At each position of the collection fiber, the heterodyne signal (reference laser and BAL) and the signal of the BAL alone are measured sequentially within a few seconds, so the spatially-resolved RF-spectrum of the BAL (see Fig.~\ref{fig:caseHighPump}d) is measured as well. An electronic fiber VOA (Thorlabs V800F controlled by a LeCroy Wavestation 3162 signal generator) is used to block the reference laser when measuring the signal of the BAL alone. Furthermore, the background signals of the reference laser and the photodetector itself were measured. For this purpose, the signal of the BAL is blocked with a mechanical shutter (Thorlabs SH05/M). The whole measurement is repeated for a second reference wavelength (shifted by about $0.5$~GHz). The complete measurement takes about $4 \sfrac{1}{2}$~minutes. Measurements are made for pump currents from $300$--$1000$~mA in steps of $50$~mA, with six measurements for somewhat different reference wavelengths at each pump current, yielding $90$~measurements in total. 

\subsection{Analysis of heterodyne measurements} \label{ssec:hetDyneAna}

\begin{figure}[tb]
\begin{center}
\includegraphics[width = 8.4 cm]{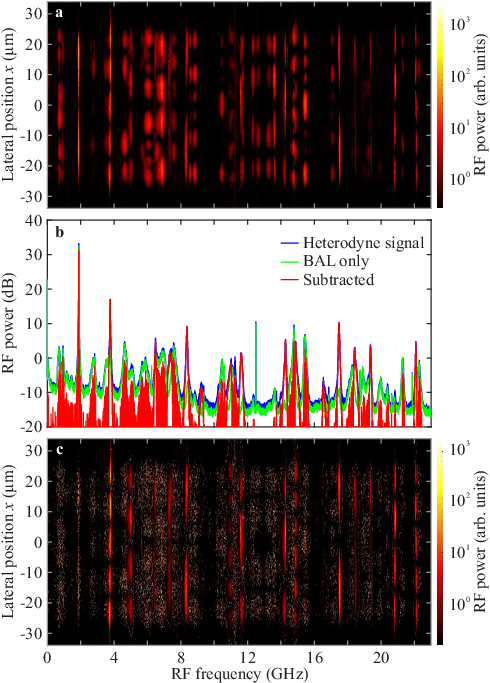}
\end{center}
\caption{\textbf{Subtraction of BAL-signal in heterodyning measurements.} \textbf{a}, RF-spectra of the heterodyne signals as function of the lateral position $x$. \textbf{b}, The RF-spectrum of the BAL alone (green) is subtracted from that of the heterodyning signal (blue) to obtain the subtracted RF-spectrum (red) analyzed in the following (example for $x = 15~\mu$m). \textbf{c}, The subtracted RF-spectra only display the beat notes between BAL and reference laser. The data shown is for a pump current of $400$~mA and a reference wavelength of $851.142$~nm.}
\label{sfig:bkgdSub}
\end{figure}

Figure~\ref{sfig:bkgdSub} shows the RF-spectra of the heterodyning signals for $400$~mA pump current, which are the modulus square of the Fourier transform of the measured time traces [see Eq.~(\ref{eq:RFspec})]. They contain both the beat notes between the lasing modes of the BAL and the reference laser and frequency components of the BAL dynamics. Hence, the RF-spectrum of the signal from the BAL alone is calculated and subtracted as shown in Fig.~\ref{sfig:bkgdSub}b. It should be emphasized that (i) the RF-power is subtracted (i.e., not the complex Fourier transform, but its modulus square), and (ii) that the subtraction is performed in linear scale, not in logarithmic scale which would mean dividing the RF-spectra. Please note that no background subtraction was performed when analyzing and plotting the RF-spectra of the BAL alone. This concerns Figs.~\ref{fig:RFspectra}, \ref{fig:caseHighPump}d and \ref{fig:caseHighPump}e, as well as Fig.~\ref{sfig:peakPropertiesAvsB}a. 

In the subtracted RF-spectrum presented in Fig.~\ref{sfig:bkgdSub}c, the only strong signals left are those of the beat notes between BAL and reference laser, though some residues of the frequency components stemming from the BAL alone still remain. From Fig.~\ref{sfig:bkgdSub}c we can deduce the spatial profiles of the lasing modes and the beat note frequencies. Since the beat note frequency $f_m$ of mode $m$ with optical frequency is $\nu_m$ is $f_m = |\nu_m - \nu_{r, 1}|$, where $\nu_{r, 1}$ is the optical frequency of the reference laser during the first measurement, we cannot obtain $\nu_m$ from Fig.~\ref{sfig:bkgdSub}c without additional information. 

\begin{figure*}[tb]
\begin{center}
\includegraphics[width = 16 cm]{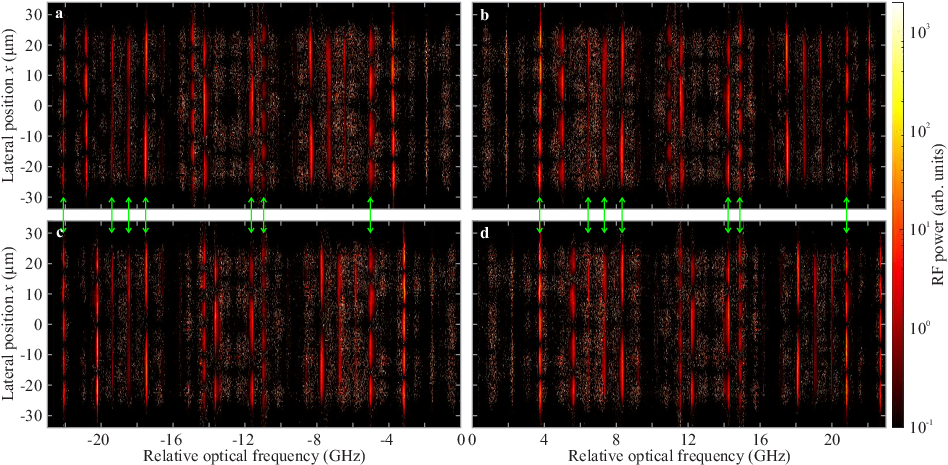}
\end{center}
\caption{\textbf{Determination of the optical frequencies of lasing modes.} Comparison of measurements with reference wavelength $\lambda_{r, 1} = 851.142$~nm (top panels \textbf{a} and \textbf{b}) and $\lambda_{r, 2} = 851.141$~nm (bottom panels \textbf{c} and \textbf{d}) at $400$~mA. The optical frequency is given with respect to $\lambda_{r, 1}$ in all panels. The spectra in the left column were calculated assuming that the lasing modes have longer wavelengths than the reference laser, and the spectra in the right column assuming they have shorter wavelengths. The green arrows indicate the modes that appear at the same relative optical frequency for both reference wavelengths.}
\label{sfig:HetDyneBlueVsRed}
\end{figure*}

A second measurement with a slightly shifted wavelength (optical frequency $\nu_{r, 2}$) of the reference laser allows us to determine the $\nu_m$ unambiguously as explained in Fig.~\ref{sfig:HetDyneBlueVsRed}. When $\nu_m < \nu_{r, 1}$ ($\nu_m > \nu_{r, 1}$), the optical frequency of mode $m$ is given by $\nu_m = \nu_{r, 1} - f_m$ ($\nu_m = \nu_{r, 1} + f_m$). Figure~\ref{sfig:HetDyneBlueVsRed} shows the resulting spectra for both cases ($\nu_m < \nu_{r, 1}$ in the left column and $\nu_m > \nu_{r, 1}$ in the right column) and for both reference wavelengths, where the reference for calculating the relative optical frequency is $\nu_{r, 1}$ in all cases. If the hypothesis that $\nu_m < \nu_{r, 1}$ ($\nu_m > \nu_{r, 1}$) is correct, then mode $m$ must appear at the same relative frequency in both spectra in the left (right) column. These modes are marked by the green arrows in each column. This procedure allows to unambiguously determine the mode frequencies $\nu_m$, even in the presence of "degenerate" modes with the same $f_m$ but different $\nu_m$. Moreover, any residues from the imperfect subtraction of the BAL signals are easily identified as such since they do not appear at the same relative frequency in either column. 

The mode frequency $\nu_{m, x}$ for each lateral position $x$ is determined as the center of mass (COM) in a $300$~MHz wide frequency window around the estimated mode frequency. In a second step the spatial average weighted with the mode amplitude $A(x)$ for each $x$ is calculated, $\nu_m = \sum_x A(x) \nu_{m, x} / \sum_x A(x)$, and finally the average over the two measurements with different reference laser wavelengths. The spatio-spectral images (Figs.~\ref{fig:caseHighPump}a and \ref{fig:caseMultiplet}a) are reconstructed by using the spatial profiles obtained from one of the two measurements with different reference wavelength, the averaged mode frequency $\nu_m$ and a Lortentzian lineshape with $100$~MHz linewidth for the sake of simplicity. The amplitudes of all modes are normalized to one to ensure good visibility of all modes. 

While the spectral resolution of the heterodyne measurement is limited only by the linewidth of the reference lasers ($< 1$~MHz), the actual measurement accuracy of the mode frequencies $\nu_m$ is not as good. Fluctuations in the data during the measurement such as slight drifts of the reference laser wavelength or small changes in the BAL state manifest as variations of $\nu_{m, x}$. These variations are typically of the order of $10$~MHz, which determines the accuracy with which $\nu_m - \nu_r$ is measured. The measurement accuracy of $\nu_m$ is limited by the absolute accuracy of $\nu_r$, which is measured by the wavelength meter with a precision of $600$~MHz. Finally, it should be noted that the spectral range of the heterodyne measurement is limited by the detection bandwidth of photodetector and oscilloscope which is $23$~GHz, yielding a spectral range of $46$~GHz ($111$~pm). 

\section{Analysis of linewidths and mode spacings}

\subsection{Linewidth determination}
The linewidths of peaks in the RF-spectra of the BAL or the heterodyne signals are determined by fitting a Lorentz function,
\begin{equation} L(f) = A \frac{\Gamma^2}{(f - f_0)^2 + \Gamma^2/4} \, , \end{equation}
where $f_0$ is the peak frequency, $\Gamma$ the full width at half maximum (FWHM), and $A$ the amplitude. In the case of spatially resolved data, a fit is made for each lateral position $x$, and the linewidth is averaged over $x$ after removing outliers and unsuitable data points. Data points are removed if (i) the fit algorithm does not converge, (ii) the fitted FWHM is unusually small or large, or (iii) the amplitude is small, which indicates low signal strength and hence unreliable fits. Criterion (ii) means that the linewidth is two time larger or smaller than the median linewidth. Criterion (iii) means that the fitted amplitude $A$ is smaller than the highest fitted amplitude divided by ten. 

The linewidth of the beat note between a lasing mode of the BAL and the reference laser is the sum of the two respective mode linewidths since the BAL and the reference laser are completely uncorrelated (see Appendix~\ref{sssec:linewidthTheory}). Since the linewidths of the BAL lasing modes are of the order of $100$~MHz, whereas the linewidth of the reference laser is $< 1$~MHz, the latter is negligible in practice, and we approximate the linewidth of the lasing mode to be equal to the linewidth of the heterodyne beat note. 

\subsection{The narrow beat note in the RF-spectra}

\begin{figure}[tb]
\begin{center}
\includegraphics[width = 8.4 cm]{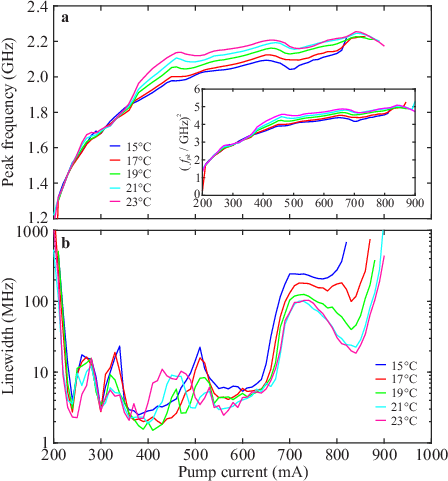}
\end{center}
\caption{\textbf{Properties of the narrow beat note.} \textbf{a},~Frequency of the narrow beat note in the RF-spectrum for five different set temperatures of the laser diode mount. The inset shows the square of the frequency as function of the pump current. \textbf{b},~Linewidth of the beat note as function of the pump current. All measurements were made with laser A.}
\label{sfig:thePeakProperties}
\end{figure}

The broad-area laser (BAL) investigated here features one particularly narrow peak in the RF-spectrum that is related to multiplets of 1st- and 2nd-order modes (see Fig.~\ref{fig:RFspectra}b). Figure~\ref{sfig:thePeakProperties} shows the frequency and linewidth of this peak as function of the pump current. Both frequency and linewidth have a complicated dependency on the pump current, and the peak frequency does not exhibit the square-root scaling expected for the relaxation oscillation frequency (see inset of Fig.~\ref{sfig:thePeakProperties}a). The linewidth of the beat note is smallest in the current region of $250$ to $650$~mA, which is the regime where the multiplets are observed. It increases rapidly for higher pump currents, indicating the loss of phase locking. Both the peak frequency and linewidth exhibit a number of local extrema and sudden slope changes. These indicate that the nonlinear dynamics creating the multiplets and the phase locking depends sensitively on the pump current, possibly in the context of gain competition between an increasing number of transverse and longitudinal modes. 

\begin{figure}[tb]
\begin{center}
\includegraphics[width = 8.4 cm]{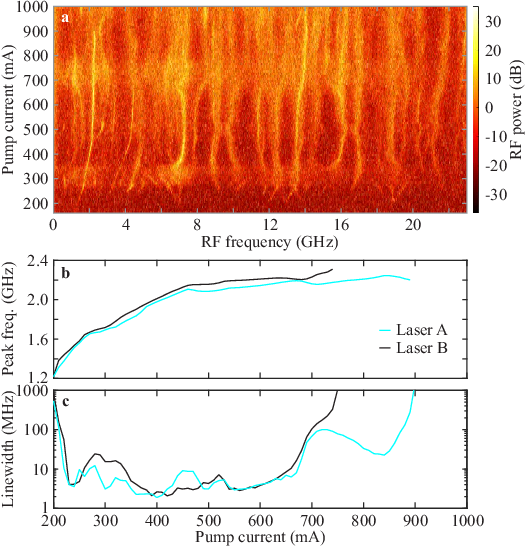}
\end{center}
\caption{\textbf{Dynamics of laser B.} \textbf{a}, RF-spectra of the total laser emission for laser B. \textbf{b},~Peak frequency and \textbf{c},~linewidth of the narrow RF-peak. The data for laser A (cyan) and laser B (black) are compared. All measurements are made at a set temperature $21^\circ$C.}
\label{sfig:peakPropertiesAvsB}
\end{figure}

Both the peak frequency and linewidth depend on the temperature. Figure~\ref{sfig:thePeakProperties} shows data for five different set temperatures of the laser diode mount ranging from $15^\circ$C to $23^\circ$C, where $21^\circ$C is the default temperature used for all other measurements. However, the effect of temperatures changes on both frequency and linewidth is relatively small. Moreover, the positions of extrema and sudden slope changes are almost the same for different temperatures. This indicates that the significant changes of the peak properties as function of the pump current are mainly due to changes in gain and mode competition rather than temperature changes induced by increased Joule heating. 

We furthermore test a second laser of the same model, called laser B in the following to distinguish it from laser A for which all other data shown in the article is measured. The RF-spectra of laser B are presented in Fig.~\ref{sfig:peakPropertiesAvsB}a, which are qualitatively similar to those of laser A (see Fig.~\ref{fig:RFspectra}a). Most importantly, laser B also exhibits a narrow peak in the RF-spectrum just like laser A. The frequency and linewidth of the narrow peak of laser B are compared to that of laser A in Figs.~\ref{sfig:peakPropertiesAvsB}b and c, showing almost quantitative agreement in the current range up to $700$~mA. Please note that the data for laser B stops around $750$~mA because the Lorentz fit used to determine the linewidth fails for higher pump currents even though the peak is still visible in the RF-spectrum till about $900$~mA. These results demonstrate that the phase-locking effect can be found for different lasers of the same model. Thus, the phase locking is reproducible and appears to be typical for certain types of broad-area semiconductor lasers. 

\subsection{Spacings and linewidths of the multiplet modes} 

\begin{figure}[tb]
\begin{center}
\includegraphics[width = 8.4 cm]{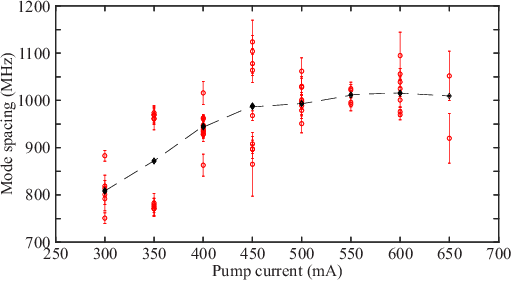}
\end{center}
\caption{\textbf{Mode spacings in the multiplets}. The NN-spacings (red) are compared to the NNN-spacings divided by two (black). Data are presented as mean +/- standard deviation for $141$ transverse positions.}
\label{sfig:NNspacings}
\end{figure}

The nearest-neighbor (NN) spacings of the modes in the multiplets are generally different from the next-nearest-neighbor (NNN) spacings divided by two as shown in Fig.~\ref{sfig:NNspacings}. This means that the modes in the multiplets are not equidistant and hence not created by four-wave mixing. It should also be noted that the NN-spacings of different multiplets at the same pump current are not necessarily the same. The NN-spacings increase with pump current, and at the same time the NNN-spacing (and thus the frequency of the narrow beat note) increases as well, so the NNN-spacing remains roughly equal to two times the NN-spacing. 

\begin{figure}[tb]
\begin{center}
\includegraphics[width = 8.4 cm]{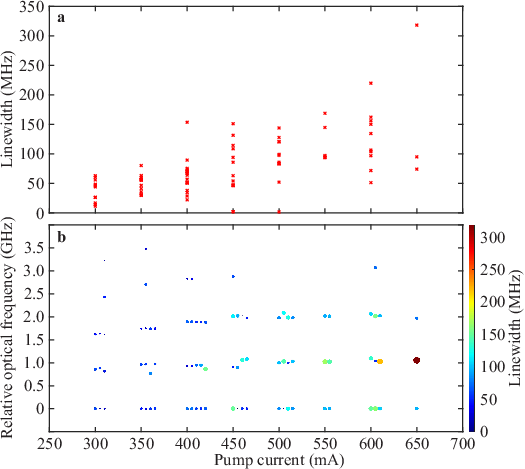}
\end{center}
\caption{\textbf{Linewidths of the multiplet modes.} \textbf{a},~Linewidths as function of the pump current. \textbf{b},~Mode linewidths organized by multiplet. The size and the color of the circles correspond to the linewidth. Different multiplets for the same pump current are offset horizontally for clarity.}
\label{sfig:multipletLinewidths}
\end{figure}

Figure~\ref{sfig:multipletLinewidths}a shows a general increase of the linewidths of the multiplet modes with pump current. It should be noted that the two data points with almost zero linewidth at $450$ and $500$~mA are artifacts in cases where the Lorentz fit fails, and the mode frequencies are not reliable either in these cases. Since the linewidth of the beat note stays roughly constant around $10$~MHz in this current regime, the linewidth reduction factor (LRF) increases as well (see Fig.~\ref{fig:caseMultiplet}e). So, increasing the pump current results in both an increase in the mode linewidth and in the strength of phase locking. 

Figure~\ref{sfig:multipletLinewidths}b shows the linewidths for each multiplet separately, where multiplets for the same pump current are horizontally offset for clarity. We systematically observe that every second mode in the multiplets is broader than its neighbors as also shown in Fig.~\ref{fig:caseMultiplet}b. 

\section{Phase fluctuations and correlations} \label{sec:phaseFlucCorr}

\subsection{Theory of linewidth and phase fluctuations} \label{sssec:linewidthTheory} 
In the following we discuss how the linewidths of lasing modes and beat notes on a photodetector are related to the phase fluctuations of the involved lasing modes, and how information on correlations due to phase locking can be extracted. Following Ref.~\cite{Gallion1984}, we neglect amplitude fluctuations of the lasing modes and consider only their phase fluctuations. 

We will use the following notation and relations: we denote the Fourier transform $\mathcal{F}$ of a complex-valued function $y(t)$ with $\tilde{y}(f) = \mathcal{F}[y(t)]$. Then the power spectral density (PSD) of $y$ is 
\begin{equation} S_y(f) = |\tilde{y}(f)|^2 = \mathcal{F}[C_y(\tau)] \end{equation}
according to the Wiener-Khinchin theorem where
\begin{equation} C_y(\tau) = \langle y^*(t) y(t + \tau) \rangle \end{equation}
is the first order autocorrelation function of $y$ and $\langle \dots \rangle$ is the time average. Furthermore, we denote the finite difference (i.e., short-term fluctuation) over time $\tau$ of a function $y$ as $\delta_\tau y(t) = y(t + \tau) - y(t)$. 

First, we consider a single lasing mode $m$ with electric field
\begin{equation} E_m(t) = A_m e^{i [\omega_m t + \phi_m(t)]} \end{equation}
where $\omega_m$ is the optical angular frequency of the mode, $A_m$ its constant amplitude, and its phase $\phi_m(t)$ follows a Wiener process. Its first-order autocorrelation function is
\begin{equation} C_{E_m}(\tau) = |A_m|^2 e^{i \omega_m \tau} \langle \exp(i \delta_\tau \phi_m) \rangle \, . \end{equation}
It is commonly assumed that the phase fluctuations $\delta_\tau \phi_m$ are a zero-mean stationary random Gaussian process \cite{Gallion1984}, that is,
\begin{equation} \label{eq:phaseFlucVsLW} \delta_\tau \phi_m(t) = \sqrt{2 \pi \Gamma_m |\tau|} z_m(t) \, , \end{equation}
where $z_m(t)$ is a Gaussian random variable with zero mean and standard deviation $\sigma_{z_m} = 1$, and $\Gamma_m$ is the linewidth of the lasing mode as shown in the following. Using the relation
\begin{equation} \label{eq:expGaussAvg} \langle \exp(i z) \rangle = \exp[-\langle z^2 \rangle / 2] = \exp[-\sigma_z^2 / 2] \end{equation}
which only holds for a Gaussian random variable $z$ with zero mean, we obtain
\begin{equation} C_{E_m}(\tau) = |A_m|^2 e^{i \omega_m \tau} e^{-\pi \Gamma |\tau|} \, . \end{equation}
The Fourier transform (FT) of the autocorrelation function yields the PSD of the electric field of mode $m$, i.e., its optical spectrum,
\begin{equation} S_{E_m}(\nu) = |A_m|^2 \frac{1}{\pi} \frac{\Gamma_m / 2}{(\nu - \nu_m)^2 + (\Gamma_m/2)^2} \, , \end{equation}
which is a Lorentz function with linewidth (FWHM) $\Gamma_m$ at optical frequency $\nu_m = \omega_m / (2 \pi)$. 

Next, we consider the beat signal of two modes $m$ and $n$ on a photodetector. The total electric field is 
\begin{equation} E_t = A_m e^{i [\omega_m t + \phi_m(t)]} + A_n e^{i [\omega_n t + \phi_n(t)]} \end{equation}
with the same assumptions for the phases of the two modes as discussed above. The photocurrent signal $j$ on the detector is $j(t) = r P(t)$ with $r$ the radiant sensitivity and $P = |E_t|^2$ the optical power. The first-order autocorrelation function of the current is $C_j(\tau) = r^2 C_{E_t}^{(2)}(\tau)$ where 
\begin{equation} C_{E_t}^{(2)}(\tau) = \langle E_t(t) E_t^*(t) E_t(t + \tau) E_t^*(t + \tau) \rangle \end{equation}
is the second-order autocorrelation function of the total electric field and we neglected shot noise. When calculating $C_{E_t}^{(2)}$, terms with an explicit time dependence drop out due to the time average, and we also remove the DC-terms that we are not interested in. This yields
\begin{equation} C_{E_t}^{(2)}(\tau) = |A_m|^2 |A_n|^2  e^{i \Delta \omega_{mn} \tau} \langle e^{i (\delta_\tau \phi_m - \delta_\tau \phi_n)} \rangle + \quad \mathrm{c.c.} \end{equation}
with $\Delta \omega_{mn} = \omega_m - \omega_n$. The further analysis depends on the properties of the random variable $\delta_\tau \phi_{mn} = \delta_\tau \phi_m - \delta_\tau \phi_n$ where $\phi_{mn} = \phi_m - \phi_n$, that is, if there are correlations between the phase fluctuations of the two modes. 

First, we consider the case of uncorrelated modes. This is for example the case in heterodyne measurements where the modes of the BAL and the reference laser are of course uncorrelated. In this case we can use that the sum of two independent Gaussian random variables $z_1$ and $z_2$ is a Gaussian random variable $z_3$ whose average is equal to the sum of the averages of $z_1$ and $z_2$ and whose variance is equal to the sum of the variances of $z_1$ and $z_2$, that is $\sigma_{z_3}^2 = \sigma_{z_1}^2 + \sigma_{z_2}^2$. Thus, the random variable $\delta_\tau \phi_{mn}$ is Gaussian and has a variance of $2 \pi |\tau| (\Gamma_m + \Gamma_n)$, and applying Eq.~(\ref{eq:expGaussAvg}) yields
\begin{equation} C_{E_t}^{(2)}(\tau) = |A_m|^2 |A_n|^2 e^{i \Delta \omega_{mn} \tau} e^{-\pi |\tau| (\Gamma_m + \Gamma_n)} \, . \end{equation}
The corresponding PSD of the photodetector signal (i.e., the RF-spectrum we measure),
\begin{equation} S_j(f) = r^2 |A_m|^2 |A_n|^2 \frac{1}{\pi} \frac{(\Gamma_m + \Gamma_n) / 2}{(f - \Delta \nu_{mn})^2 + (\Gamma_m + \Gamma_n)^2 / 4} \, , \end{equation}
is also a Lorentz function, where $\Delta \nu_{mn} = (\omega_m - \omega_n) / (2 \pi)$. Hence the linewidth of a heterodyne beat note that we measure is the linewidth of the BAL mode plus the linewidth of the reference laser. Since the latter is $< 1$~MHz while the BAL modes have typically linewidths of $50$~MHz and more, the contribution from the reference laser can be neglected in practice. 

Next, we consider the case of modes with correlated phase fluctuations, such as the NNN-mode pairs in the multiplets which are phase-locked. In that case the variance of $\delta_\tau \phi_{mn}$ is
\begin{equation} \begin{array}{rcl} \var(\delta_\tau \phi_{mn}) & = & \var(\delta_\tau \phi_m) + \var(\delta_\tau \phi_n) \\ & & - 2 \cov(\delta_\tau \phi_m, \delta_\tau \phi_n) \\ \\
 & = & 2 \pi |\tau| (\Gamma_m + \Gamma_n) \\ & & - 4 \pi |\tau| \sqrt{\Gamma_m \Gamma_n} \cov(z_m, z_n) \end{array} \end{equation}
where $z_{m, n}$ are Gaussian random variables with unit variance [see Eq.~(\ref{eq:phaseFlucVsLW})] and $\cov$ denotes the covariance. It should be noted that 
\begin{equation} \cov(z_m, z_n) = \rho_{z_m, z_n} = \rho_{\delta_\tau \phi_m, \delta_\tau \phi_n} =: \rho_{m,n} \, , \end{equation}
where $\rho_{x, y}$ is the Pearson correlation coefficient of two functions $x$ and $y$, and $\rho_{m, n}$ is the phase fluctuation correlation (PFC) of modes $m$ and $n$. 

At this point it should be noted that the sum of two \textit{correlated} random Gaussian variables is not necessarily a Gaussian variable itself. Thus, Eq.~(\ref{eq:expGaussAvg}) is not guaranteed to be applicable and the lineshape of the beat note not necessarily a Lorentz function. The calculation of the term $\langle e^{i \delta_\tau \phi_{mn}} \rangle$ requires the evaluation of all moments of $\delta_\tau \phi_{mn}$ which include higher-order correlations between $\delta_\tau \phi_m$ and $\delta_\tau \phi_n$. Hence, a careful analysis of the lineshape of the beat note of two phase-locked modes could yield information about higher-order phase fluctuation correlations, not just the linear correlation $\rho_{m, n}$. In practice, the narrow beat note in the RF-spectrum is well fitted by a Lorentz function, and measuring the lineshape in greater detail to observe possible deviations from a Lorentz profile is challenging. 

\begin{figure}[tb]
\begin{center}
\includegraphics[width = 8.4 cm]{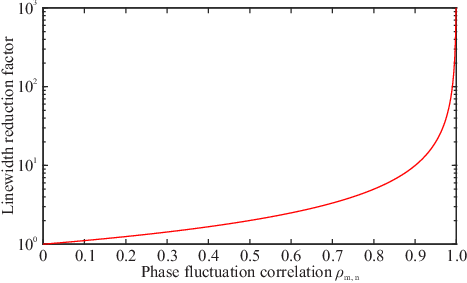}
\end{center}
\caption{\textbf{Linewidth reduction factor.} The LRF for equal mode linewidths, Eq.~(\ref{eq:LRFsimple}), is shown as function of the phase fluctuation correlation $\rho_{m, n}$.}
\label{sfig:LRFvsCorr}
\end{figure}

For the sake of simplicity we will hence assume that $\delta_\tau \phi_{mn}$ is Gaussian and apply Eq.~(\ref{eq:expGaussAvg}), yielding a Lorentzian beat note with linewidth
\begin{equation} \Gamma_b = \frac{\var(\delta_\tau \phi_{mn})}{2 \pi |\tau|} = \Gamma_m + \Gamma_n -2 \sqrt{\Gamma_m \Gamma_n} \, \rho_{m, n} \, . \end{equation}
We define the linewidth reduction factor (LRF) as
\begin{equation} \LRF = \frac{\Gamma_m + \Gamma_n}{\Gamma_b} = \left[ 1 - 2 \frac{\sqrt{\Gamma_m \Gamma_n}}{\Gamma_m + \Gamma_n} \rho_{m, n} \right]^{-1} \, , \end{equation}
which simplifies to
\begin{equation} \label{eq:LRFsimple} \LRF = \frac{1}{1 - \rho_{m, n}} \end{equation}
in the limit of equal mode linewidths, $\Gamma_m \approx \Gamma_n$. Equation~(\ref{eq:LRFsimple}) is plotted in Fig.~\ref{sfig:LRFvsCorr} and shows that a LRF of $\LRF = 10$ corresponds to a phase fluctuation correlation of $\rho_{m, n} = 0.9$. 

Finally, we consider how the phase fluctuation correlations of two laser modes $m$ and $n$ can be determined from the heterodyne signals, that is, their beat notes with the reference laser $r$. The power of the beat notes between mode $m$ and the reference laser is
\begin{equation} \label{eq:beatNoteTime} P_{mr}(t) = 2 \mathrm{Re}[A_m A_r \cos(\Delta \omega_{mr} t + \phi_{mr})] \, , \end{equation}
where we removed the DC-terms which our photodetector filters out. We apply the Hilbert transform $\mathcal{H}$ to obtain the analytic signal
\begin{equation} P_{mr} + i \mathcal{H}[P_{mr}] = 2 A_m A_r \exp\{ i (\Delta \omega_{mr} t + \phi_{mr}) \} \end{equation}
from which we extract the instantaneous amplitude, 
\begin{equation} \label{eq:anaSignAmpExtract} A_{mr}(t) = |P_{mr} + i \mathcal{H}[P_{mr}]| = 2 A_m A_r \, , \end{equation}
and the instantaneous phase,
\begin{equation} \theta_{mr}(t) = \arg\{ P_{mr} + i \mathcal{H}[P_{mr}] \} = \Delta \omega_{mr} t + \phi_{mr}(t) \, . \end{equation}
The relative phase of mode $m$ and the reference laser is obtained as
\begin{equation} \label{eq:anaSignPhaseExtract} \phi_{mr}(t) = \theta_{mr}(t) - t \left \langle \frac{d \theta_{mr}}{dt} \right \rangle_t \end{equation}
where we estimate $\Delta \omega_{mr}$ as the average of the numerical derivative of $\theta_{mr}$. From that we obtain the relative phase fluctuations
\begin{equation} \delta_\tau \phi_{mr}(t) = \sqrt{2 \pi |\tau|} \left( \sqrt{\Gamma_m} z_m(t) - \sqrt{\Gamma_r} z_r(t) \right) =: \delta \phi_m \end{equation}
that we denote as $\delta \phi_m$ in the following. In practice, $\tau = 20$~ps is the sampling step of the oscilloscope, and $\delta \phi_m$ is dominated by the phase fluctuations of the lasing mode since its linewidth $\Gamma_m$ is much larger than $\Gamma_r$. 

The same procedure is applied to the beat note of mode $n$ and the reference laser, yielding $\delta \phi_n$. Then we calculate the Pearson correlation coefficient of the two relative phase fluctuations,
\begin{equation} \rho_{\delta \phi_m, \delta \phi_n} = \frac{\mathcal{E}[ \delta \phi_m \delta \phi_n ]}{\sigma_{\delta \phi_m} \sigma_{\delta \phi_m}}  \, , \end{equation}
where $\mathcal{E}(y)$ denotes the expectation value and $\sigma_y$ the standard deviation of a function or variable $y$. Since the two lasing modes are not correlated with the reference laser, we find $\sigma_{\delta \phi_{m, n}} = \sqrt{2 \pi |\tau| (\Gamma_{m, n} + \Gamma_r)}$. The product of the relative phase fluctuations in the nominator is
\begin{equation} \begin{array}{rcl} \delta \phi_m \delta \phi_n & = & 2 \pi |\tau| \left( \Gamma_r z_r^2 + \sqrt{\Gamma_m \Gamma_n} z_m z_n \right. \\ \\ & & \left. - \sqrt{\Gamma_m \Gamma_r} z_m z_r - \sqrt{\Gamma_n \Gamma_r} z_n z_r \right) \, . \end{array} \end{equation}
When we calculate its expectation value, the last two terms vanish since lasing modes and reference laser are uncorrelated, whereas $\mathcal{E}[z_m z_n] = \rho_{m, n}$, which yields
\begin{equation} \rho_{\delta \phi_m, \delta \phi_n} = \frac{\Gamma_r + \sqrt{\Gamma_m \Gamma_n} \rho_{m, n}}{\sqrt{\Gamma_m + \Gamma_r} \sqrt{\Gamma_n + \Gamma_r}} \, . \end{equation}
Again, we simplify by assuming that $\Gamma_m \approx \Gamma_n$ to obtain
\begin{equation} \rho_{\delta \phi_m, \delta \phi_n} = \frac{1}{1 + \gamma} + \frac{\gamma}{1 + \gamma} \rho_{m, n} \end{equation}
with $\gamma = \Gamma_m / \Gamma_r$. Thus, the phase fluctuation correlation of the two beat notes has an offset of $1 / (1 + \gamma)$ stemming from the phase fluctuations of the reference laser which are present in both beat notes. However, this effect is again negligible for the lasing mode linewidths considered here, so we find $\rho_{\delta \phi_m, \delta \phi_n} \approx \rho_{m, n}$ in good approximation. 

So in principle we can determine the PFC $\rho_{m, n}$ of two lasing modes from the correlation of the phase fluctuations of the two heterodyne signals. In practice, the beat note signals of the two modes must be extracted from the measured heterodyne time traces using a relatively narrow frequency filter due to the presence of many other beat notes, and this filter modifies the phase fluctuations and reduces their correlations as discussed in the next section. 

We conclude this section with a remark why one should consider the correlations between the phase fluctuations $\delta_\tau \phi_{m, n}(t)$ and not that between the phases $\phi_{m, n}(t)$ themselves. Following Eq.~(\ref{eq:phaseFlucVsLW}), the phase $\phi_m(t)$ is a sum over many uncorrelated realizations of a Gaussian random variable $z_m(t)$, in other words, the phase performs a one-dimensional random walk. However, it is well known that two random walks can show high (positive or negative) correlations even if the Gaussian variables generating the two random walks are completely uncorrelated \cite{Granger1974, Phillips1986}. These correlations are spurious in the sense that there is no mathematical or physical meaning attached to them. The spurious correlations persist even for arbitrarily long time series \cite{Phillips1986}. This phenomenon can be explained as follows: a typical random walk with $M$ steps will have a total displacement proportional to $\sqrt{M}$ either in positive or negative direction. Consequently, two random walks going in the same (opposite) direction will show a significant correlation (anti-correlation), whereas low correlations are rarely found because only few random walks have a total displacement close to zero. Hence, for random-walk like time series it is good practice to consider the fluctuations around their mean behavior instead when searching for possible correlations \cite{Granger1974}. 

\subsection{Simulations} \label{sssec:phaseCorrSims} 
In this section we present simulations of a heterodyning signal in order to demonstrate how the phase fluctuations can be extracted from it and to understand the effect of applying a frequency filter. We numerically calculate a heterodyning beat note $P_{mr}(t)$ as in Eq.~(\ref{eq:beatNoteTime}) with a sampling step of $\tau = 20$~ps where we choose constant mode amplitudes $A_m = A_r = 1$, a beat note frequency $\Delta \nu_{mr} = 2$~GHz, and linewidths of $\Gamma_m = 100$~MHz and $\Gamma_r = 1$~MHz for the lasing mode and the reference laser, respectively. The phases $\phi_{m}$ and $\phi_r$ of lasing mode and reference laser are calculated as in Eq.~(\ref{eq:phaseFlucVsLW}). The beat note $P_{mr}(t)$ is a sinusoidal signal with phase modulated by the phase fluctuations of the two lasing modes. 

\begin{figure}[tb]
\begin{center}
\includegraphics[width = 8.4 cm]{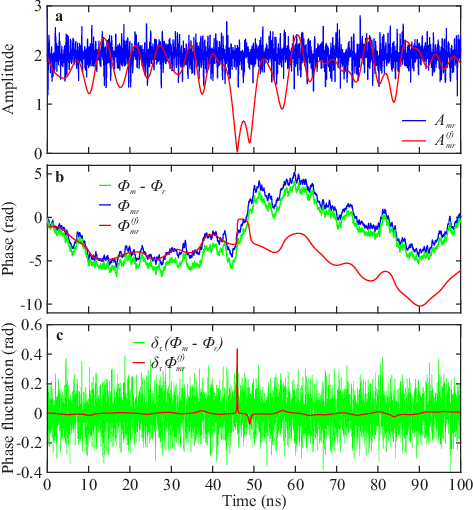}
\end{center}
\caption{\textbf{Analysis of simulated beat note signal}. \textbf{a},~Amplitude of the analytic signal of the simulated beat note without filter ($A_{mr}$, blue) and with filter ($A_{mr}^{(f)}$, red). \textbf{b},~Phase extracted from the simulated beat note without filter ($\phi_{mr}$, blue) and with filter ($\phi_{mr}^{(f)}$, red) as well as the original phase ($\phi_m - \phi_r$, green). \textbf{c},~Original phase fluctuations [$\delta_\tau(\phi_m - \phi_r)$, green] and phase fluctuations extracted with filter ($\delta_\tau \phi_{mr}^{(f)}$, red).}
\label{sfig:phaseFlucSim}
\end{figure}

First, we apply the Hilbert transform directly to $P_{mr}(t)$ and extract the amplitude $A_{mr}(t)$ and the phase $\phi_{mr}(t)$ of the beat note as in Eqs.~(\ref{eq:anaSignAmpExtract})--(\ref{eq:anaSignPhaseExtract}). The amplitude shown as blue line in Fig.~\ref{sfig:phaseFlucSim} exhibits small fluctuations around $2$ even though it is expected to be constant and equal to $2$. Figure~\ref{sfig:phaseFlucSim}b compares the phase $\phi_{mr}(t)$ (blue) extracted from the analytic signal according to Eq.~(\ref{eq:anaSignPhaseExtract}) to the original phase $\phi_m - \phi_r$ (green) used to calculate the beat note, which agree well. This demonstrates that the procedure outlined above allows us to determine the phase fluctuations of the lasing modes from the beat notes, even though there are small deviations between extracted amplitude and phase compared to the simulated ones. 

However, in practice the measured heterodyne signals contain several beat notes as well as contributions from the BAL dynamics, hence a relatively narrow filter must be applied around the individual beat notes in order to separate them from the other signals. To test the effect of filtering, we apply a $300$~MHz wide filter [cf.\ Eq.~(\ref{eq:filter})] to the simulated beat note $P_{mr}(t)$. Then we calculate the analytic signal of the filtered time trace $P_{mr}^{(f)}$ and extract amplitude $A_{mr}^{(f)}$ and phase $\phi_{mr}^{(f)}$ as above, which are shown in red in Fig.~\ref{sfig:phaseFlucSim}. The amplitude of the filtered signal shows severe fluctuations, including dips going to almost zero. These fluctuations are artifacts induced by the filter: in the event of very fast phase changes, for example at $46$~ns in Fig.~\ref{sfig:phaseFlucSim}b, the instantaneous frequency of the beat note can leave the frequency range of the filter, thus reducing the amplitude of the filtered signal. This unfortunately means that our heterodyne analysis does not allow us to determine a possible fast amplitude dynamics of the lasing modes since any genuine fluctuations of the amplitude will be obscured by the spurious amplitude fluctuations due to the filter. 

The phase (Fig.~\ref{sfig:phaseFlucSim}b) is also affected by the filter. First, it is smoothed because very fast phase fluctuations are filtered out. Hence the phase fluctuations of the filtered signal $\delta_\tau \phi_{mr}^{(f)}$ (red line in Fig.~\ref{sfig:phaseFlucSim}c) have a smaller variance than the original ones (green line) and are also smooth, not randomly fluctuating like the original ones. Second, strong dips in the amplitude lead to spurious fluctuations of the phase (phase slips) like at $46$~ns where $\phi_{mr}^{(f)}$ no longer follows the original phase. These phase slips manifest in strong extrema of the phase fluctuations as seen in Fig.~\ref{sfig:phaseFlucSim}c at $46$~ns. Once the amplitude has recovered (from $55$~ns on), the filtered phase once again follows the original one, but with an offset. In order to remove these spurious phase fluctuations induced by the filter, $4$~ns long windows around local minima of the amplitude are removed from the phase fluctuation time traces before calculating their correlation (see Appendix~\ref{sssec:phaseFlucExpAna}). 

\begin{figure}[tb]
\begin{center}
\includegraphics[width = 8.4 cm]{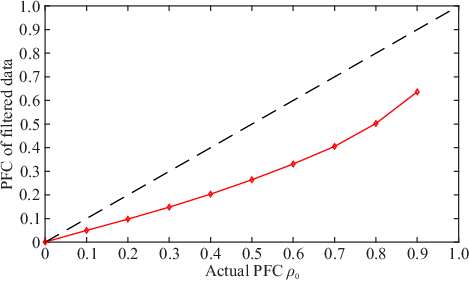}
\end{center}
\caption{\textbf{Phase fluctuation correlations $\rho_{m, n}^{(f)}$ of filtered beat notes.} The black dashed lines indicates $\rho_0$ as visual reference.}
\label{sfig:phaseCorrFilterRed}
\end{figure}

Finally, we investigate how the data analysis procedure affects the correlation measurement. To this end, we calculate the beat note $P_{nr}$ of another lasing mode $n$ with the reference laser. Lasing modes $n$ and $m$ have the same linewidth of $100$~MHz, but we now introduce a correlation $\rho_0$ of their phase fluctuations. The phase fluctuations of mode $n$ are calculated from the random variable $z_n$ [see Eq.~(\ref{eq:phaseFlucVsLW})] given by
\begin{equation} z_n = \rho_0 \, z_m + \sqrt{1 - \rho_0^2} \, z_3 \, , \end{equation}
where $z_m$ and $z_3$ are independent Gaussian random variables with unit variance, and $z_m$ is the random variable determining the phase fluctuations of mode $m$. By construction, $\rho_{z_m, z_n} = \rho_0$. 

We calculate the Pearson correlation coefficient $\rho_{m, n}^{(f)}$ of the filtered phase fluctuations $\delta_\tau \phi_{mr}^{(f)}$ and $\delta_\tau \phi_{nr}^{(f)}$, after additionally removing the parts near local minima of the amplitude as described above. The resulting correlations $\rho_{m, n}^{(f)}$ are compared to the actual correlations $\rho_0$ of the simulated phase fluctuations in Fig.~\ref{sfig:phaseCorrFilterRed}. We find that for small $\rho_0$, the extracted PFCs are $\rho_{m, n}^{(f)} \simeq \rho_0 / 2$ and remain significantly lower even for high $\rho_0$. In conclusion, the analysis procedure used to measure the PFCs yields values that are systematically lower than the actual correlations, which partially explains the relatively low PFCs observed experimentally. Nonetheless, our simulations demonstrate that we obtain qualitatively correct information about possible phase correlations between different lasing modes. 

\subsection{Analysis of experimental data} \label{sssec:phaseFlucExpAna} 

First, the heterodyne signals of individual lasing modes are extracted by applying a Fourier filter. The Fourier transform $\mathcal{F}$ of the heterodyne time trace $U(t, x)$ measured at lateral position $x$ is calculated,
\begin{equation} \tilde{U}(f, x) = \mathcal{F}\{U(t, x)\} \, , \end{equation}
where $f$ is the RF-frequency. Since $\tilde{U}(f, x)$ contains contributions from may different beat notes as well as the dynamics of the BAL itself, a filter over a frequency interval $[f_1, f_2]$ around the beat note frequency $f_m = |\nu_m - \nu_r|$ is applied, where $\nu_m$ is the optical frequency of the lasing mode $m$ and $\nu_r$ that of the reference laser. So the filtered time-domain beat note signal of mode $m$ is
\begin{equation} \label{eq:filter} U_m^{(f)}(t, x) = \mathcal{F}^{-1}\{ \tilde{U}(f, x) [B_{f_1, f_2}(f) + B_{-f_2, -f_1}(f)] \} \, , \end{equation}
where $B_{f_1, f_2}(t)$ is the boxcar function with unit amplitude over the interval $[f_1, f_2]$. It should be noted that $U_m^{(f)}(t, x)$ contains possible signals from the BAL itself in that frequency range which cannot be subtracted in a meaningful way. 

It should be noted that some lasing modes are "degenerate" in the RF-spectrum, that is, they have almost the same beat note frequency $f_m$ because their optical frequencies are symmetrical with respect to the reference laser frequency. Hence, it is impossible to separate them with a filter as explained in the previous section, and they are left out of the analysis since they would give spurious results. 

The filter interval for mode $m$ is given by
\begin{equation} f_{1, 2} = f_m \mp 3 \Gamma_m / 2 \, , \end{equation}
where $\Gamma_m$ is the full width at half maximum (FWHM) of the beat note averaged over $x$, $\Gamma_m = \langle \Gamma_{m, x} \rangle_x$. Here, the FWHM $\Gamma_{m, x}$ at position $x$ is obtained from the RF-spectrum of the heterodyne signal (smoothed with a $f_\mathrm{avg} = 10$~MHz moving average),
\begin{equation} \label{eq:HDspec} P_{hd}(f, x) = \mathcal{M}\{ |\tilde{U}(f, x)|^2 \} \, , \end{equation}
by determining the $-3$~dB points around the peak at $f_m$. Positions for which the FWHM could not be properly determined due to, e.g., low signal strength, are ignored when calculating $\Gamma_m$. A total width of $3 \Gamma_m$ for the filter is used as a compromise between the requirements of having a large bandwidth and removing noise and interfering signals as much as possible. For a Lorentzian line shape, the filter thus contains $79.5\%$ of the total RF-power of the beat note signal. In practice, the filter width is of the order of a few $100$~MHz. 

The beat note $U_m^{(f)}(t, x)$ is a sinusoidal signal with a mean oscillation frequency $f_m$ whose phase is modulated by the phase fluctuations of lasing mode $m$ of the BAL and the reference laser. We apply the Hilbert transform to obtain the corresponding analytic signal and decompose it into the instantaneous amplitude $A_{mr}^{(f)}(t, x)$ and instantaneous phase $\theta_{mr}^{(f)}(t, x)$ such that $U_m^{(f)} = \mathrm{Re}[A_{mr}^{(f)} \exp(i \theta_{mr}^{(f)})]$. The average slope of $\theta_{mr}^{(f)}$ corresponds to the beat frequency $f_m$, and hence we calculate 
\begin{equation} \phi_{mr}^{(f)} = (\theta_{mr}^{(f)} - 2 \pi t f_m) \mathrm{sgn}(\nu_m - \nu_r) \end{equation}
where $\mathrm{sgn}$ is the sign function. The phase $\phi_{mr}^{(f)}$ is the difference between the instantaneous phases of the lasing mode $m$ and the reference laser. In the following we consider the phase fluctuations 
\begin{equation} \delta_\tau \phi_{mr}^{(f)}(t, x) = \phi_{mr}^{(f)}(t + \tau, x) - \phi_{mr}^{(f)}(t, x) \end{equation}
where $\tau = 20$~ps is the sampling step of the oscilloscope. 

The phase fluctuations $\delta_\tau \phi_{mr}^{(f)}(t, x)$ are the difference between the phase fluctuations of the lasing mode and that of the reference laser, where the latter are negligible compared to the former since the linewidth of the reference laser is very small compared to that of the lasing modes in our case (see Appendix~\ref{sssec:linewidthTheory}). It should be noted that the standard deviation of $\delta_\tau \phi_{mr}^{(f)}(t, x)$ is significantly reduced compared to that of the actual phase fluctuations due to the filter that is applied (see Fig.~\ref{sfig:phaseFlucSim}). 

Furthermore, very fast phase fluctuations lead to high instantaneous frequencies that cause the signal to (partially) leave the frequency domain of the boxcar filter that is applied (see Fig.~\ref{sfig:phaseFlucSim}). In such cases, $\delta_\tau \phi_{mr}^{(f)}$ shows spurious values, and the instantaneous amplitude $A_{mr}^{(f)}$ exhibits a strong dip. To remove these filter artifacts, $4$~ns long pieces of $\delta_\tau \phi_{mr}^{(f)}$ are removed around pronounced dips in the amplitude. Finally, the Pearson correlation coefficient $\rho_{m, n}^{(f)}(x)$ between these reduced versions of $\delta_\tau \phi_{m}^{(f)}(t, x)$ and $\delta_\tau \phi_{n}^{(f)}(t, x)$ is calculated. As a consequence of the unavoidable filtering process, the measured phase fluctuation correlations $\rho_{m, n}^{(f)}(x)$ are somewhat lower than the actual ones (see Fig.~\ref{sfig:phaseCorrFilterRed}). 

\begin{figure}[tb]
\begin{center}
\includegraphics[width = 8.4 cm]{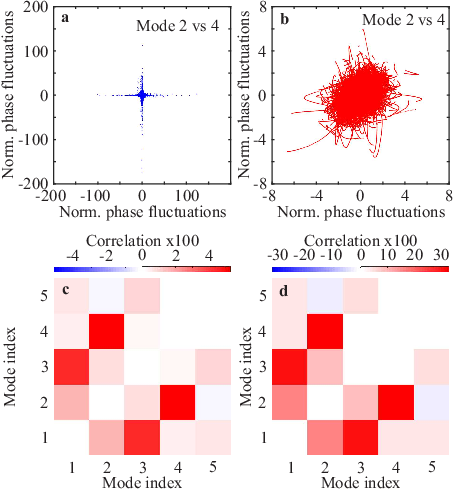}
\end{center}
\caption{\textbf{Effect of removing phase slips on correlations.} The data is for five modes at $350$~mA for lateral position $x = 15~\mu$m. These are the quintuplet modes shown in Fig.~\ref{fig:caseMultiplet}. \textbf{a},~Correlation diagram for the phase fluctuations $\delta_\tau \phi_{mr}^{(f)}$ of modes $2$ and $4$ before removal of the phase slips and \textbf{b}~after their removal. The phase fluctuations are normalized by their standard deviation. The correlation coefficients are $0.05$ and $0.33$, respectively. \textbf{c},~Correlation matrix of the five modes for $x = 15~\mu$m before removal of phase slips and \textbf{d},~after their removal. Please note the different color scales.}
\label{sfig:PFCcalcEffectOfReduction}
\end{figure}

\begin{figure*}[p]
\begin{center}
\includegraphics[width = 14 cm]{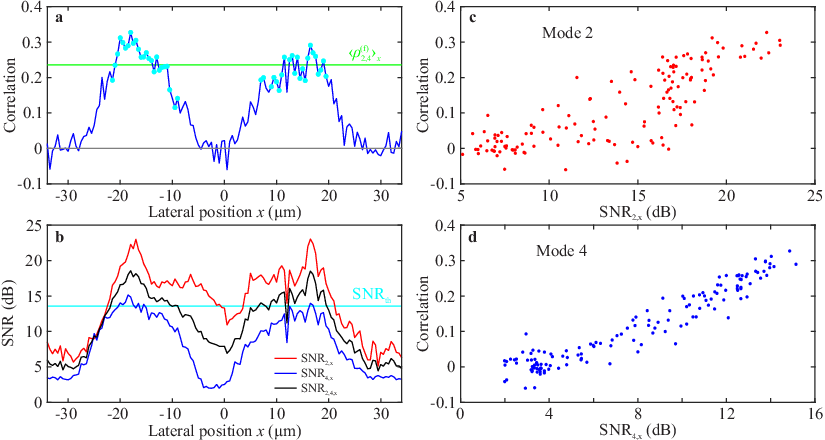}
\end{center}
\caption{\textbf{Phase fluctuation correlations and SNR, first example.} Data at $350$~mA for mode $2$ at $-11.52$~GHz and mode $4$ at $-9.78$~GHz (cf.\ Fig.~\ref{sfig:PFCcalcEffectOfReduction}) is shown. \textbf{a},~PFC as function of lateral position $x$ (blue). The data points with a joint SNR above the threshold $\SNR_{th}$ are indicated by cyan dots, and their average $\langle \rho_{2, 4}^{(f)} \rangle_x = 0.24$ is indicated by the green line. \textbf{b},~SNR of modes $2$ (red) and $4$ (blue) as well as their joint SNR (black) as function of $x$. The SNR threshold $\SNR_{th}$ is indicated by the cyan line. \textbf{c},~Correlation diagram between the PFC and the SNR of mode $2$ (Pearson correlation of $0.81$). \textbf{d},~Correlation diagram between the PFC and the SNR of mode $4$ (Pearson correlation of $0.96$).}
\label{sfig:PFCvsX-SNR1}
\end{figure*}

\begin{figure*}[p]
\begin{center}
\includegraphics[width = 14 cm]{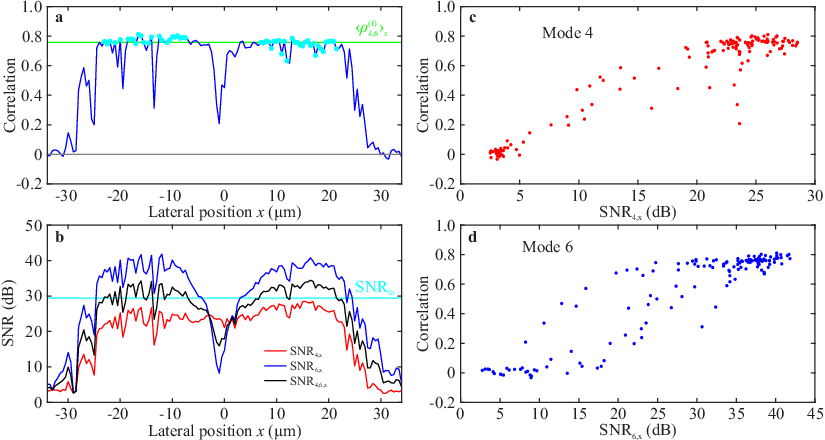}
\end{center}
\caption{\textbf{Phase fluctuation correlations and SNR, second example.} Data at $300$~mA for mode $4$ at $14.09$~GHz and mode $6$ at $15.70$~GHz is shown. \textbf{a},~PFC as function of lateral position $x$ (blue). The data points with a joint SNR above the threshold $\SNR_{th}$ are indicated by cyan dots, and their average $\langle \rho_{4, 6}^{(f)} \rangle_x = 0.76$ is indicated by the green line. \textbf{b},~SNR of modes $4$ (red) and $6$ (blue) as well as their joint SNR (black) as function of $x$. The SNR threshold $\SNR_{th}$ is indicated by the cyan line. \textbf{c},~Correlation diagram between the PFC and the SNR of mode $4$ (Pearson correlation of $0.95$). \textbf{d},~Correlation diagram between the PFC and the SNR of mode $6$ (Pearson correlation of $0.92$).}
\label{sfig:PFCvsX-SNR2}
\end{figure*}

Figure~\ref{sfig:PFCcalcEffectOfReduction} shows how strongly the phase slips that occur as artifacts at local minima of the amplitude affect the measured PFCs. The correlation diagram of two modes of a quintuplet at $350$~mA in Fig.~\ref{sfig:PFCcalcEffectOfReduction}a has a cross shape which is due to the very strong phase fluctuations that appear during the phase slips. Consequently, the correlation coefficient is only $0.05$. By removing $4$~ns long windows around the phase slips as described above, these spurious fluctuations are removed, and the correlation diagram becomes an elliptic point cloud with a correlation coefficient of $0.33$ as shown in Fig.~\ref{sfig:PFCcalcEffectOfReduction}b. The correlation matrices of the quintuplet modes before and after removal of the phase slips are shown in Figs.~\ref{sfig:PFCcalcEffectOfReduction}c and d, respectively. The correlations of almost all mode pairs increase drastically when removing the phase slips, and all correlation data shown in the main text is calculated after their removal. Please note that the correlation values in Fig.~\ref{sfig:PFCcalcEffectOfReduction}d are different from those in Fig.~\ref{fig:phaseCorr}a because the latter includes a spatial average of the PFCs as explained in the following. 

Moreover, detector noise and signals from the dynamics of the BAL that fall within the boxcar filter bandwidth significantly reduce the measured PFCs as well (see Fig.~\ref{sfig:PFCvsX-SNR1}). The signal-to-noise ratio of the beat note for mode $m$ at position $x$ is given by 
\begin{equation} \SNR_{m, x} = \max_{f} \left( \frac{P_{hd}(f, x)}{P_\mathrm{BAL}(f, x)} B_{f_1, f_2}(f) \right) \, \end{equation}
where the RF-spectrum $P_\mathrm{BAL}$ of the BAL alone is calculated from its time trace $U_\mathrm{BAL}(t, x)$ in the same way as $P_{hd}$. The joint SNR of the mode pair $(m, n)$ is defined as their geometric mean (or arithmetic mean in logarithmic scale), 
\begin{equation} \label{eq:SNRmean} \SNR_{m, n, x} = \sqrt{ \SNR_{m, x} \cdot \SNR_{n, x} } \, , \end{equation}
and the abscissa of Figs.~\ref{fig:phaseCorr}b and c is simply $\max \limits_x (\SNR_{m, n, x})$. 

Figures~\ref{sfig:PFCvsX-SNR1}a and b show the PFC and the signal-to-noise ratio (SNR) of two modes with $q = 1$ and $2$ as function of the lateral position $x$. Not surprisingly, the SNR and in consequence the PFC is low at positions where the modes have low intensities. Figures~\ref{sfig:PFCvsX-SNR1}c and d show the correlation diagrams of the PFC and the SNR of the individual modes, which demonstrate how strongly the measured PFCs depend on a good SNR. Hence, we only consider the data points with the best joint SNR, that is, 
\begin{equation} \label{eq:SNRthres} \SNR_{m, n, x} > \SNR_{th} = \max_x (\SNR_{m, n, x}) - 5 \, \mathrm{dB} \, . \end{equation}
We calculate $\langle \rho_{m, n}^{(f)} \rangle_x$ as the average of $\rho_{m, n}^{(f)}(x)$ over the data points fulfilling this criterion. It is the quantity $\langle \rho_{m, n}^{(f)} \rangle_x$ which is shown in Fig.~\ref{fig:phaseCorr}. The SNR threshold is indicated by the cyan line in Fig.~\ref{sfig:PFCvsX-SNR1}b, and the corresponding data points are indicated by cyan dots in Fig.~\ref{sfig:PFCvsX-SNR1}a. The average of these data points is $\langle \rho_{2, 4}^{(f)} \rangle_x = 0.24 \pm 0.05$ (green line in Fig.~\ref{sfig:PFCvsX-SNR1}a). 

The example in Fig.~\ref{sfig:PFCvsX-SNR1} is typical for mode pairs with fairly strong correlation and decent SNR. Figure~\ref{sfig:PFCvsX-SNR2} shows the example of a NNN-mode pair at $300$~mA which has the highest correlation that was found in the measurement data, $\langle \rho_{4, 6}^{(f)} \rangle_x = 0.76 \pm 0.03$. The sum of the linewidths of these lasing modes is $40.7$~MHz compared to a linewidth of $9.2$~MHz of the beat note, which yields a linewidth reduction factor of $\LRF = 4.4$. For comparison, the measured PFC of $0.76$ corresponds to $\LRF = 4.14$ according to Eq.~(\ref{eq:LRFsimple}), which agrees almost quantitatively. We see in Figs.~\ref{sfig:PFCvsX-SNR2}c and d that the PFC saturates with increasing SNR, and it seems that this saturation indicates when the correct value of the PFC is reached. However, it should be noted that this example has an unusually low LRF; in most cases the $\LRF \geq 10$ (see Fig.~\ref{fig:caseMultiplet}e), but measuring PFCs above $0.9$ appears to be out of reach with the current setup due to insufficient SNR. 

\subsection{Comparison to other phase measurement techniques}
A commonly used technique to detect phase locking and measuring the relative phases of modes in frequency combs is shifted wave interference Fourier transform spectroscopy (SWIFTS) \cite{Hugi2012, Burghoff2014, Singleton2018}. However, we cannot use this technique for characterizing our BAL because SWIFTS requires a reference signal with the frequency of the mode spacing, but our laser exhibits many different frequency spacings that are not known \textit{a priori}. Our heterodyne technique is hence more versatile than SWIFTS, in particular for highly multimode lasers and in cases where not all mode pairs are phase-locked. On the other hand, its disadvantages are the need for a reference laser which also limits the precision of frequency and phase measurements, a spectral range that is limited to two times the detection bandwidth ($2 \times 23$~GHz in our case), and the need for a good SNR. Since measuring very high phase correlations correctly is difficult with heterodyning as shown above, our technique is more adapted to detecting weak or partial phase locking like in our BAL, whereas SWIFTS is adapted for strong phase locking as in the case of frequency combs. Hence heterodyning could be useful, for example, to investigate the transition to mode locking before full coherence between modes is reached. 

In Ref.~\cite{Cappelli2019} heterodyning was used to investigate mid-infrared and terahertz frequency combs. The main differences of Ref.~\cite{Cappelli2019} to our work are that a frequency comb was used as reference laser instead of a single-mode laser as in our case, and both the reference laser and the laser under test were stabilized with a phase-locked loop, whereas we do not have any external stabilization. Another difference to SWIFTS and the heterodyne technique in Ref.~\cite{Cappelli2019} is that they measure the long-term stability of mode locking, whereas we can consider phase fluctuations on very short time scales, which is more suitable for lasers without additional long-term stabilization. 

\section{Measurements with multimode-fiber photodetectors} \label{sec:MMF}
Measuring the dynamics of multi-transverse mode lasers with fiber-coupled high-speed photodetectors presents a number of experimental problems that may impact the measurement results. First, the coupling efficiency of different transverse laser modes into a single-mode fiber (SMF) or multimode fiber (MMF) is not necessarily the same. In particular high-order transverse modes may have lower coupling efficiencies. Consequently, the relative amplitudes of different transverse modes in the signal transmitted by the fiber can be different from the actual spectrum of the laser. Second, the configuration of a fiber, that is, how it is laid out and coiled on the optical table, affects the polarization state of the transmitted signal and, in the case of MMFs, the partitioning of the signal in different transverse fiber modes because small defects and birefringence induced by mechanical stress couple and mix the various fiber modes during propagation. Third, different transverse laser modes couple differently to the various transverse fiber modes. Since the overlap of different transverse modes in the signal on the photodetector affects the generated photocurrent, the coupling of the laser into the fiber and the configuration of the fiber can affect the measurement results, in particular the amplitudes of beat notes between different laser modes in the RF-spectra. In addition, for heterodyne measurements the polarization overlap of the signals from the BAL and the reference laser will affect the amplitude of the heterodyne beat notes. 

In order to clarify the practical importance of these issues, we perform measurements of the time traces of the BAL alone as well as heterodyne measurements with different couplings into the fiber and different configurations of the fibers transmitting the laser signals to the photodetector. These experiments allow to quantify how reproducible the RF-spectra calculated from the measured time traces are, how strongly the results are affected by the experimental conditions, and which aspects of the setup have a significant impact. 

\subsection{RF-spectra of total laser emission} \label{ssec:RFspecTotalVsMMF}
The setup to measure the total laser emission and the MMF are described in Appendix~\ref{ssec:timeDomMeas}. Different aspects of the setup (fiber configuration, coupling into the fiber) are varied, and measurements for the following eight setup variants are performed:
\begin{itemize}
\item Coupling into the MMF is optimized (\textit{MMF coupling 1}), and for the same configuration of the fiber (\textit{MMF config 1}), three consecutive measurements (\textit{no 1--3}) are made (i.e., the setup was not changed during these three measurements).
\item The coupling into the MMF remains unchanged (\textit{MMF coupling 1}), but the configuration of the MMF is changed two times (\textit{MMF config 2, 3}).
\item The coupling efficiency into the MMF is reduced by moving the MMF-mount in the fiber coupler slightly in the horizontal direction (\textit{MMF coupling 2}), i.e., the transverse dimension of the BAL. The coupling efficiency is reduced even further (\textit{MMF coupling 3}) in the same way. The configuration of the MMF was not changed during these two measurements.
\item A SMF (Thorlabs P1-780A-FC-2, $5~\mu$m mode diameter, NA = $0.13$, $2$~m long) is installed (\textit{SMF coupling}) instead of the MMF and coupled to a single-mode detector (Newport 1484-A, $15$~kHz to $22$~GHz).
\end{itemize}
When changing the fiber configuration, the fibers are laid out on the optical table in very different ways while avoiding bends with a small radius. These deliberate changes in the fiber configuration are much more significant than changes that occur involuntarily during the experiments. 

Each time the coupling into the fiber is changed, the LI-curve of the power in the fiber is measured without ND-filters in order to determine the quality of the coupling. Then, ND-filters are added to reduce the power at $1000$~mA to $1$~mW before measuring the time traces. For each setup variant the time traces in the range of $160$--$1000$~mA are measured (duration of about $9$~minutes per measurement). The RF-spectra presented in the following are calculated from the measured time traces with a $10$~MHz moving average [see Eq.~(\ref{eq:RFspec})] in order to reduce noise fluctuations as much as possible to facilitate the direct comparison of different RF-spectra (in contrast to the $2$~MHz moving average used for the results in Fig.~\ref{fig:RFspectra}). 

\begin{figure}[tb]
\begin{center}
\includegraphics[width = 8.4 cm]{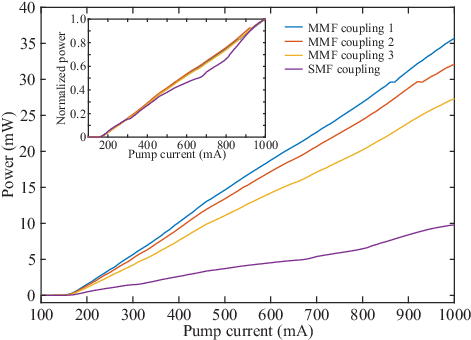}
\end{center}
\caption{\textbf{LI-curves of power coupled into a fiber.} The LI-curves are for three different coupling configurations with a multimode fiber and for coupling into a single-mode fiber. Inset: LI-curves normalized to $1$ at $1000$~mA.}
\label{sfig:LIcurveFibers}
\end{figure}

Figure~\ref{sfig:LIcurveFibers} shows the measured LI-curves for the four different coupling configurations (\textit{MMF coupling 1,2,3} and \textit{SMF coupling}). The maximal power in the MMF is deliberately reduced by about $10\%$ for \textit{MMF coupling 2} and by $24\%$ for \textit{MMF coupling 3} with respect to optimal coupling (\textit{MMF coupling 1}). When coupling into a SMF, only about $27\%$ of the power possible for the MMF were attained at $1000$~mA. The normalized LI-curves in the inset of Fig.~\ref{sfig:LIcurveFibers} show that the three LI-curves for the MMF are quite, though not perfectly linear above threshold. In contrast, the LI-curve for the SMF exhibits sublinear behavior with a significant dip in the range of $500$--$900$~mA. We attribute this to lower coupling efficiencies for some of the higher-order transverse modes of the BAL, which means that the signal in the SMF is biased and does not represent the actual distribution of power among different transverse modes. The quite good linearity of the three LI-curves for the MMF, in contrast, indicates that there is no significant transverse mode bias when coupling into a MMF, even when the overall coupling efficiency is somewhat reduced. Hence, we consider measurements with a MMF more reliable and adequate than measurements with a SMF when investigating broad-area lasers with multiple transverse modes. 

\begin{figure*}[tb]
\begin{center}
\includegraphics[width = 16 cm]{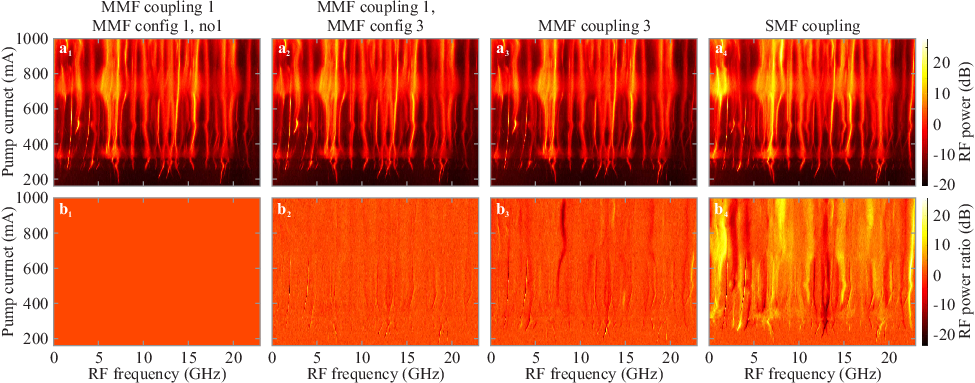}
\end{center}
\caption{\textbf{RF-spectra of total laser emission for different configurations of fiber and coupling.} \textbf{a}, RF-spectra obtained from measured time traces. \textbf{b}, Difference $\Delta P_{RF}$ with respect to the reference spectrum in panel a$_1$. From left to right: \textit{MMF coupling 1} with \textit{MMF config 1} and measurement \textit{no 1} (a$_1$, b$_1$), \textit{MMF coupling 1} with \textit{MMF config 3} (a$_2$, b$_2$), \textit{MMF coupling 3} (a$_3$, b$_3$), and \textit{SMF coupling} (a$_4$, b$_4$).}
\label{sfig:RFspectraComparison}
\end{figure*}

\begin{figure}[tb]
\begin{center}
\includegraphics[width = 8.4 cm]{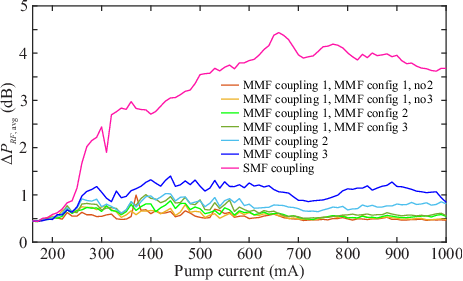}
\end{center}
\caption{\textbf{Average deviations from reference RF-spectrum.} The quantity $\Delta P_{RF, \mathrm{avg}}^{(i)}$ from Eq.~(\ref{eq:DeltaPrfAvg}) is shown for the different fiber and coupling configurations.}
\label{sfig:RFspecMeanDiff}
\end{figure}

Figure~\ref{sfig:RFspectraComparison}a shows four examples of RF-spectra for different setup variants. To compare different RF-spectra $P_{RF}^{(i)}(f, I_f)$, where $f$ is the RF-frequency, $I_f$ the pump current and $i$ the setup variant, we calculate their ratio (i.e., difference in logarithmic scale) as
\begin{equation} \Delta P_{RF}^{(i)}(f, I_f) = P_{RF}^{(i)}(f, I_f) / P_{RF}^{(\mathrm{ref})}(f, I_f) \end{equation}
where we use the very first measurement (\textit{coupling 1, MMF config 1, no 1}) as reference spectrum $P_{RF}^{(\mathrm{ref})}$. These are shown in Fig.~\ref{sfig:RFspectraComparison}b. The three measurements for the same coupling and MMF configuration are so similar that only the first one is shown in Fig.~\ref{sfig:RFspectraComparison}a$_1$. Even when changing the configuration of the MMF (Figs.~\ref{sfig:RFspectraComparison}a$_2$ and b$_2$), only rather small deviations are visible. It should be noted that differences visible in Fig.~\ref{sfig:RFspectraComparison}b$_2$ are mostly due to tiny changes of the frequencies of certain frequency components caused by thermal drift than actual changes of their amplitude. When degrading the coupling (Figs.~\ref{sfig:RFspectraComparison}a$_3$ and b$_3$), more and stronger deviations appear. However, some frequency components in the RF-spectrum are affected more strongly than others, and the overall deviations remain relatively small. When coupling into a SMF (Figs.~\ref{sfig:RFspectraComparison}a$_4$ and b$_4$), in contrast, most of the frequency components show significant changes of their amplitudes compared to the case of coupling into a MMF, and even a few frequency components not visible for MMF coupling appear. 

For a quantitative comparison of the RF-spectra, Fig.~\ref{sfig:RFspecMeanDiff} shows the frequency-averaged deviations from the reference RF-spectrum which are calculated as
\begin{equation} \label{eq:DeltaPrfAvg} \Delta P_{RF, \mathrm{avg}}^{(i)}(I_f) = \langle 10 \log_{10} \{ |\Delta P_{RF}^{(i)}(f, I_f)| \} \rangle_f \end{equation}
i.e., the average is performed in decibel scale. The red and orange curves (repetition measurements of the original setup) show average deviations of only about $0.5$~dB over the whole current range. This is the level of deviations that occurs naturally when no aspect of the setup is changed. The two green curves (change of the MMF configuration without changing the coupling) barely deviate from the red and orange curves, which means that changing the MMF configuration has practically no effect on the RF-spectra. The two blue curves (change of coupling efficiency into the MMF) show increasing deviations for most pump current regimes as the coupling is degraded, demonstrating that the coupling into the MMF is an important experimental parameter. Nonetheless, the resulting changes are on average not higher than $1.5$~dB, and Fig.~\ref{sfig:RFspectraComparison}b$_3$ shows that only a few of the frequency components are significantly affected. The magenta curve for coupling into a SMF shows much higher deviations than all the others, typically $> 3$~dB for most of the pump current range, as expected from Figs.~\ref{sfig:RFspectraComparison}a$_4$ and b$_4$. 

We draw several conclusions from the experimental observations discussed above. First, the insignificant dependency of the measured signal on the configuration of the MMF shows that mixing and coupling of different fiber modes during propagation does not affect the measured signals for MMFs that are only a few meter in length. However, the impact of the MMF configuration may be stronger for long fibers. Second, we conclude from the LI-curves in Fig.~\ref{sfig:LIcurveFibers} that we can couple all transverse modes of the BAL with more or less equal efficiency into a MMF, and hence the signal measured with the MMF-coupled photodetector should be representative of the actual fluctuations of the total laser power. Third, degrading the coupling into the MMF affects the measured RF-spectra, probably due to unequal coupling efficiency for different transverse laser modes. But the resulting transverse mode bias is not very large as indicated by the relatively minor changes of the LI-curves and RF-spectra in Figs.~\ref{sfig:LIcurveFibers} and \ref{sfig:RFspecMeanDiff}, respectively. From the insignificant effect of the MMF configuration and the rather small effect of imperfect coupling we conclude that measurements of RF-spectra with MMF-coupled photodetectors are reliable and a good representation of the actual fluctuations of the BAL emission. Even though we cannot completely exclude the possibility that imperfect coupling of the transverse laser modes in the MMF affects the amplitudes of beat notes, the good linearity of the LI-curves in Fig.~\ref{sfig:LIcurveFibers} leads us to believe that this is a minor problem at best when coupling is carefully optimized. 

Using a SMF, in contrast, leads to a significant transverse mode bias which strongly affects the amplitudes of the frequency components and can even add new ones. We hence do not recommend the use of SMF-coupled photodetectors when investigating lasers with multiple transverse modes. It is interesting to note that all the frequency components visible in the RF-spectra measured with MMF-coupling are also found in the RF-spectra for SMF-coupling, which implies that mixing of transverse fiber modes in the MMF does not artificially create beat notes since these would otherwise not be present for measurements with a SMF. 

We would like to add that similar test measurements were made with a broad-area VCSEL \cite{Bittner2022}, leading to the same conclusions, though the test results were not published. Furthermore, we performed test measurements (not shown here) with a free-space high-speed photodetector (Newport 818-BB-51, $12.5$~GHz bandwidth, $40~\mu$m detector diameter). While using a free-space photodetector avoids possible problems with the signal propagation in a fiber, focusing the entire beam on a very small detector is at least as challenging as coupling it into a fiber, and measurement artifacts can arise when the beam is only partially focused on the detector. Hence, measurements with a free-space photodetector are not necessarily more reliable than those using fiber-coupled photodetectors. 

\subsection{Heterodyne signals with spatial resolution}
The setup for spatially-resolved heterodyne measurements is shown in Fig.~\ref{sfig:setupHetDyne}. The following measurements are performed for $1000$~mA pump current, and the collection MMF is placed at a local maximum of the intensity distribution in the image plane and kept fixed. The power of the BAL (reference laser) on the photodetector is set to $0.5$~mW ($0.75$~mW) as usual. Measurements for different configurations of the MMF transmitting the signal of the BAL and the SMF transmitting the reference laser (see Fig.~\ref{sfig:setupHetDyne}b) are performed. The RF-spectra are calculated from the measured time traces with a moving average of $10$~MHz. 

\begin{figure}[tb]
\begin{center}
\includegraphics[width = 8.4 cm]{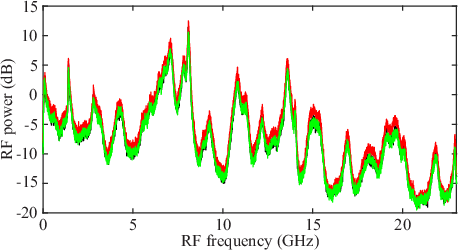}
\end{center}
\caption{\textbf{Spatially-resolved RF-spectra for different multimode fiber configurations.} The RF-spectra are measured for $1000$~mA pump current with the MMF at one of the maxima of the transverse intensity distribution for three different configurations of the MMF.}
\label{sfig:RFspecSpatialMMFcoils}
\end{figure}

First, we perform measurements of the BAL alone without the reference laser, similar to Appendix~\ref{ssec:RFspecTotalVsMMF}:
\begin{itemize}
\item Three consecutive measurements without changing the MMF configuration or any other aspect of the setup
\item Two more measurements with different configurations of the MMF
\end{itemize}
A comparison of the RF-spectra for the three different MMF configurations is shown in Fig.~\ref{sfig:RFspecSpatialMMFcoils}. There is no discernible difference between the three RF-spectra except that one of them has a slightly higher overall amplitude, which we attribute to a small increase of the coupling efficiency when the collection MMF was moved. The three measurements without setup variation (not shown) show no differences either. This confirms the results and conclusions in Appendix~\ref{ssec:RFspecTotalVsMMF} concerning the negligible effect of the MMF configuration on measurements of the BAL signal. 

\begin{figure}[tb]
\begin{center}
\includegraphics[width = 8.4 cm]{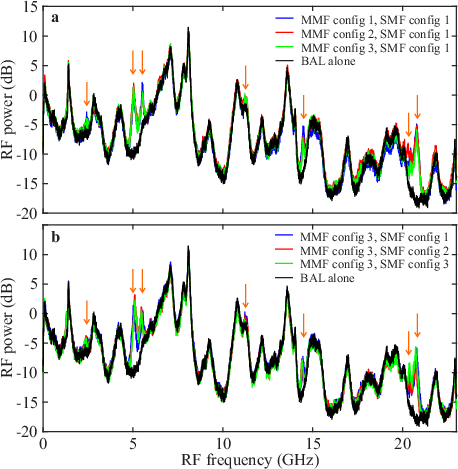}
\end{center}
\caption{\textbf{Heterodyne signals at $\mathbf{1000}$~mA for different fiber configurations.} \textbf{a}, Heterodyne signals for three different configurations of the MMF while the SMF is kept unchanged. \textbf{b}, Heterodyne signals for three different configurations of the SMF while the MMF is kept unchanged. The RF-spectrum of the BAL alone is shown in black as reference in both panels, and the heterodyne beat notes are indicated by the orange arrows. The wavelength of the reference laser is $850.292$~nm.}
\label{sfig:HetDyneFiberConfig}
\end{figure}

Next, we perform heterodyne measurements with different configurations of the MMF and SMF:
\begin{itemize}
\item Three different configurations of the MMF (\textit{MMF config 1, 2, 3}) without changing the SMF (\textit{SMF config 1}) 
\item Two more measurements with \textit{MMF config 3} and different SMF configurations (\textit{SMF config 2, 3}) 
\end{itemize}
The impact of the MMF configuration on the heterodyne signals is shown in Fig.~\ref{sfig:HetDyneFiberConfig}a, and that of the SMF configuration in Fig.~\ref{sfig:HetDyneFiberConfig}b. In both cases we see small variations of the amplitudes of the heterodyne beat notes, but only by a few dB (typically well below $5$~dB). Small variations of the peak positions observed for some case are attributed to variations of the reference laser wavelength. There seems to be no qualitative difference between the effect of changing the MMF or SMF configuration. 

We conclude that changing the fiber configurations affects the polarization state of the BAL and reference laser signals on the photodetector, which leads to small variations of the amplitudes of the heterodyne beat notes as the overlap of the polarization states of the two signals changes. Since the effect of moving the SMF, which has two polarization states but only a single transverse mode, is about as big as the effect of moving the MMF, we believe that possible changes of the transverse mode composition of the BAL signal in the MMF and the spatial overlap of the signals on the photodetector are not important. Because both signals are elliptically polarized after passing through several meters of fiber, variations of their polarization state lead to relatively small changes of the polarization state overlap which is unlikely to vanish completely. It should be noted that during normal spatially-resolved heterodyne measurements, the MMF configuration barely changes since the end of the MMF in the image plane is moved by no more than a few millimeters, and the SMF configuration is not changed at all. So while we have an uncertainty of several dB for the absolute amplitudes of the different BAL modes, the spatial profiles are not affected and reliably measured. 

%

\end{document}